\documentclass[%
superscriptaddress,
 amsmath,amssymb,
 aps,
 prb,
 twocolumn
]{revtex4-2}

\usepackage{graphicx}
\usepackage{dcolumn}
\usepackage{bm}
\usepackage{xcolor}
\usepackage[caption=false]{subfig}
\usepackage{tikz}
\usepackage{xfrac}
\usepackage{blindtext}
\usepackage{times}
\usepackage{comment}

\usepackage[normalem]{ulem}

\definecolor{darkblue}{HTML}{004D6B}
\definecolor{darkred}{HTML}{8c1515}
\definecolor{darkgreen}{HTML}{006400}
\usepackage{hyperref}
\hypersetup{
	colorlinks=true,
	urlcolor=darkblue,
	citecolor=darkred,
	linkcolor=darkblue,
	breaklinks
}

\begin{document}


\title{Transmon platform for quantum computing challenged by chaotic fluctuations}

\author{Christoph Berke}
\affiliation{Institute for Theoretical Physics, University of Cologne, 50937 Cologne, Germany}

\author{Evangelos Varvelis}
\affiliation{Institute for Quantum Information, RWTH Aachen University, 52056 Aachen, Germany}
\affiliation{J\"ulich-Aachen Research Alliance (JARA), Fundamentals of Future Information Technologies, 52425 J\"ulich, Germany}

\author{Simon Trebst}
\affiliation{Institute for Theoretical Physics, University of Cologne, 50937 Cologne, Germany}

\author{Alexander Altland}
\affiliation{Institute for Theoretical Physics, University of Cologne, 50937 Cologne, Germany}

\author{David P. DiVincenzo}
\affiliation{Institute for Quantum Information, RWTH Aachen University, 52056 Aachen, Germany}
\affiliation{J\"ulich-Aachen Research Alliance (JARA), Fundamentals of Future Information Technologies, 52425 J\"ulich, Germany}
\affiliation{Peter Gr\"unberg Institute, Theoretical Nanoelectronics, Forschungszentrum J\"ulich, 52425 J\"ulich, Germany}%


\date{\today}

\begin{abstract}
From the perspective of many body physics, the transmon qubit architectures currently developed for quantum computing  are systems of coupled nonlinear quantum resonators. A certain amount of intentional frequency detuning (`disorder') is crucially required to protect individual qubit states against the destabilizing effects of nonlinear resonator coupling. Here we investigate the stability of this variant of a many-body localized (MBL) phase for system parameters relevant to current quantum processors developed by the IBM, Delft, and Google consortia, considering the cases of natural or engineered disorder.  Applying three independent diagnostics of localization theory --- a Kullback-Leibler analysis of spectral statistics, statistics of many-body wave functions (inverse participation ratios), and a Walsh transform of the many-body spectrum --- we find that some of these computing platforms are dangerously close to a phase of uncontrollable chaotic fluctuations.    
\end{abstract}

\maketitle



When subject to strong external disorder, wave functions of
many body quantum systems may  localize in states defined by (but not in trivial ways)
the eigenstates of the disordering operators. 
A standard paradigm in this context  is the spin-\sfrac{1}{2} Heisenberg chain in a  random $z$-axis magnetic field. 
Here, the disorder basis comprises the `physical' $p$-qubits 
\footnote{In the MBL literature these are called $p$-bits and $l$-bits; we modify the notation to connect to modern quantum-information usage.}
defined by the spin states, different due to spin exchange from the eigenbasis of `localized' $l$-qubits 
\cite{PhysRevB.90.174202,PhysRevLett.111.127201}. 
The latter are stationary but remain non-trivially correlated, including in the deeply localized phase. 

Although it may seem paradoxical at first sight, intentional
`disordering' and \emph{many body localization} (MBL) in the above sense are  a
vitally important resource in the most advanced quantum computing (QC) platform
available to date, the superconducting transmon qubit array processor. Physically,
the transmon array is a system of coupled nonlinear quantum oscillators. At the low
energies relevant to QC the system becomes  equivalent to the negative $U$ Bose
Hubbard model. Site occupations $0$ and $1$ define the transmon $p$-qubit states,
known as `bare qubits' in QC language.  Randomization of the
individual qubit energies maintains the integrity of these states  in the presence of
the finite inter-transmon coupling required for computing functionality. This
coupling makes the eigen-$l$-qubits of the system different from the  $p$-qubits.
Considerable efforts are invested in the characterization and control of the induced
correlations, known as ZZ couplings in the parlance of the QC
community~\cite{PhysRevLett.125.200504}.

Connections between MBL and superconducting qubits have been considered earlier
\cite{orell_probing_2019,Throckmorton2020}, but mainly with a focus on applications
of qubit arrays as quantum simulators of the bosonic MBL transition
\cite{Throckmorton2020}. Surprisingly, however, the obvious reverse question has not
been been asked systematically so far: What bearings may qubit isolation by disorder
have on QC functionality? Reliance on strong disorder localization is a Faustian
approach inasmuch as it invites the presence of \emph{quantum chaos}, which is an
arch-enemy of quantum device control of any kind. Lowering the strength of disorder
brings one closer to the MBL-to-chaos transition, heralded by a growth of $l$-qubit
correlations as early indicators for the proximity of the uncontrollable chaotic
phase. Since the key requirement of QC, the execution of gate operations, requires
on-demand rapid growth of entanglement between $l$-qubits, it is imperative that some
definite amount of coupling is present. A crucial question that we confront,
therefore, is under which circumstances the necessary levels of coupling keep us
outside the chaotic zone.

While this question does not have an easy overall answer, one general statement can
be made with confidence: True to its Faustian nature, the invitation of disorder into
the platform can be renegotiated, but not revoked. For instance, 
in its road map for future devices, IBM aims to replace random variations of qubit frequencies by a
precision engineered frequency alternation, e.g., $\ldots$ -A-B-A-B- $\ldots$.
While this pattern efficiently blocks resonances between neighboring qubits,
\emph{next} nearest neighbors are now approximately degenerate. In a nonlinear
system, such degeneracies are  potent triggers for  instabilities; the only
way to control, or `localize' (qubit) states is with degeneracy lifting and
translational symmetry breaking -- in short, with the retention of some frequency
disorder in the A and B sets.

With this general situation in mind, the purpose of this paper is twofold~\footnote{We may incidentally remark that this work started from a project study~\cite{borner_simon-dominik_classical_2020}  on \emph{classical} chaos in the transmon system. Classically, the transmon Hamiltonian Eq.~\eqref{trans1} describes a system of coupled mathematical pendula of mass $m=1/8E_C$ and gravitational acceleration $g=8E_C E_J$ (in units where $\hbar=1$ and $\ell=1$ (pendulum length)). Nonlinearly  coupled pendula generally show a transition from integrable motion at low energies to hard chaos at high energies. (There are  desktop gimmicks with just two coupled masses demonstrating the phenomenon.) The principal observation of the project was that already the classical two transmon Hamiltonian showed tendencies to chaos when excited to sufficiently high energies.  
The generalization to ten coupled oscillators made the situation worse, with Lyapunov
 exponents signaling uncontrollable dynamics for energies matching those of QC applications with $0$ and $1$ qubit states,  and at time scales way below  typical coherence times.}.
 In its first part, we apply state-of-the-art diagnostic tools of MBL theory to investigate the role of disorder
 in transmon qubit arrays. We consider realistic models of qubit arrays employed in
 the remarkable experimental efforts by the groups of Delft~\cite{PhysRevApplied.8.034021},
 Google~\cite{GoogleQuantumSupremacy}, IBM~\cite{8936946}, and others, assuming that
 device imperfections lead to random variations of individual qubit frequencies.
 Within this framework, we describe the diminishing localization of many-$l$-qubit
 wave functions, and  the growth of $l$-qubit correlations, upon \emph{lowering}
 disorder. Considering small instances of multi-transmon systems, we  find that the
phase boundary between MBL and quantum chaos indeed may  come dangerously
 close to the parameter ranges of current experiments.  We also find that increasing
 the coordination number of the transmon lattice, as necessary for 2D connected
 transmon networks, increases many-body delocalization and the incipient
 chaos of the dynamics.

In the second part of the paper, we apply this diagnostic machinery to address the
 question of whether precision engineering may be employed to ultimately realize
 `clean' devices. Considering the above mentioned IBM alternating sequence as a
 case study, we find that it may indeed be operated at low values of randomness.
 However, for the reasons indicated above, residual frequency variations remain
  required to  safeguard the stability of the device; further
 purification will not merely lead to little further improvement, but will actually
 be {\em detrimental} to its operation.  Importantly, the diagnostic framework
 developed in the paper may be applied to predict levels of randomness which lead to
 optimal localization of quantum information for given parameters characterizing the
 clean device.

The general conclusion of this work is that further progress towards larger QCs will
be dependent on skirting the dangerous attributes of chaotic parts of the parameter
space. We know from experience with general many body systems 
that tendencies to long
range correlations and delocalization increase with increasing two-dimensional system
connectivity \cite{PhysRevX.5.031033}. On this basis,  the monitors provided by many body localization theory
may become an essential resource in the perfection of future transmon based
information devices.

\section*{Results}

{\bf Overview. } In what follows,
we introduce our principal object of study: a transmon array modeled with realistic qubit parameters. Anticipating the importance of nonlinearities, we  use effective `low-energy Hamilitonians' solely to gain intuition of the underlying physics, but perform all subsequent computations avoiding any such approximations.   We then introduce diagnostics inspired by MBL theory 
and apply them to detect signatures of chaos. 
We discuss enhanced tendencies to instability emerging in two-dimensional geometries, and 
address the question  of whether the ideal of a stable and `perfectly clean' array can be reached by advanced  qubit engineering. 


{\bf Transmon array Hamiltonian. }
\label{sec:TransmonArray}
Our study begins with the well-established minimal model for interacting transmon qubits~\cite{koch_charge-insensitive_2007,Gambetta}:
\begin{equation}
H = 4E_C\sum_i n_i^2 - \sum_i E_{J_i}\cos\phi_i + T\sum_{\langle i,j \rangle}n_{i}n_{j} \,.
\label{trans1}
\end{equation}
Here, $n_i$ is the Cooper-pair number operator of transmon $i$, conjugate to its
superconducting phase $\phi_i$. The transmon charging energy $E_C$ is determined by
the capacitance of the metal body of the transmon, and is easily fixed at a desired
constant, 
typically about $E_C=250$~MHz ($h=1$). The Josephson energy
$E_J$ is proportional to the critical current of the junction. Except in very recent work, it has been difficult to
fix this constant reproducibly to better than a few percent. However, typical values
lie in the vicinity of around
$12.5$~GHz, much larger than the charging energy.
Finally, electrical coupling between the transmons, often via a capacitance, produces
the charge coupling $T n_i n_j$. The  coefficient
$T$ has varied over a substantial range in 15 years of experiments \cite{blais_circuit_2020}; $T$ values beyond
50~MHz are possible, but $T<E_C$ is a fundamental constraint. 
Current experiments are often in the range $T=3$--$5$~MHz,
making $T$ the smallest energy scale in the problem.

\begin{figure} 
    \includegraphics[scale = 1]{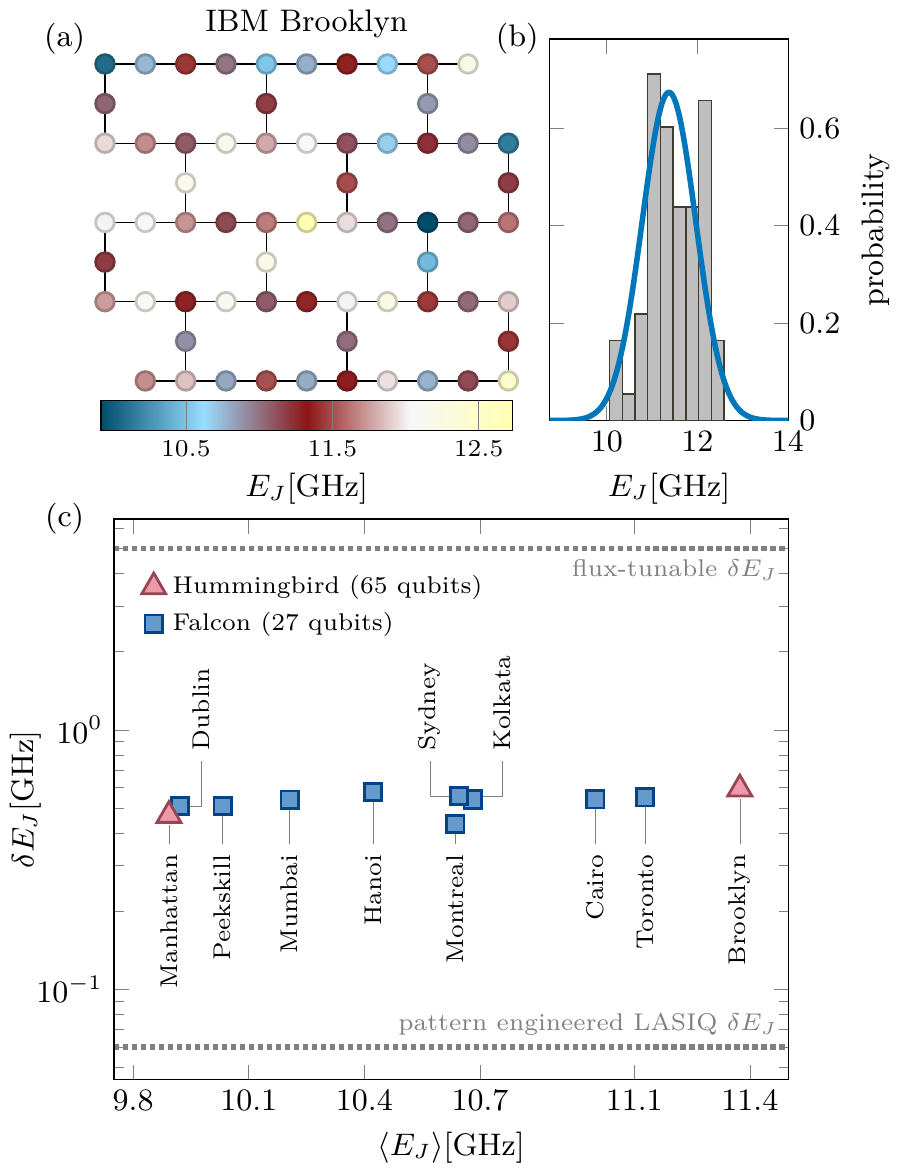}
    \caption{{\bf Experimental parameters of recent IBM transmon arrays.} 
    (a) Layout of the 65-qubit transmon array ``Brooklyn", currently available in IBM's quantum cloud \cite{IBMQuantumCloud}, 
    in a heavy hexagon geometry .
    	  The coloring of the qubits indicates the variation of Josephson energies $E_J$  
	  which is largely uncorrelated in space. 
    (b) Spread of the $E_J$ plotted for the ``Brooklyn" chip, consistent with a Gaussian distribution (solid line).
    	 Similar levels of disorder and distributions are found in all transmon devices available in IBM's quantum cloud. 
    (c) Variance of the measured Josephson energies, $\delta E_J$,  for nine realizations of the 27-qubit ``Falcon" design, 
    	and two realizations of the 65-qubit ``Hummingbird" design. 
    While the mean varies unsystematically from device to device, the variance remains very consistent, 
    setting the parameter favored in our ``scheme {\sc a}" study below. 
    ``Scheme {\sc b}" cases in other labs have a much larger spread as indicated by the ``flux tunable" level in the figure. 
    Recent proposals of using high precision laser-annealing \cite{hertzberg2020laserannealing} 
    as a pattern engineering approach \cite{zhang2020highfidelity},  
    discussed towards the end of the manuscript, 
    aim for a significant reduction of the $E_J$ variance; 
    such pattern-tuned transmon arrays have so far not appeared in any cloud device.}
    \label{Fig:real-IBM-chips}
\end{figure} 

\begin{figure*}[t] 
   \centering 
   \hspace{-1.5cm}
    \includegraphics{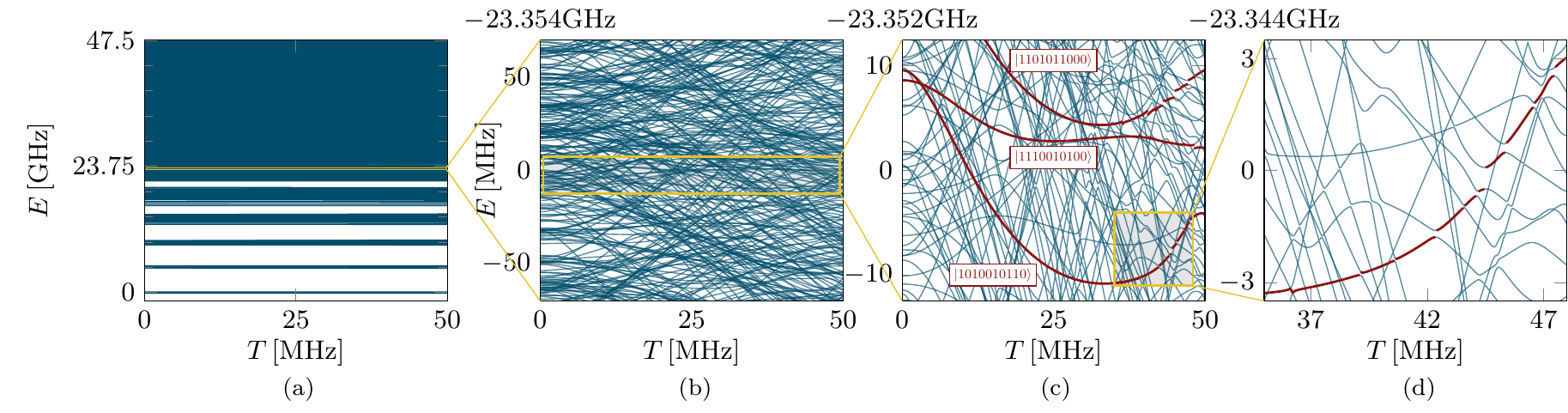}
    \caption{\textbf{Energy spectrum of a coupled transmon array.}
    		Illustrated are the energy levels $E_\alpha(T)$ of Hamiltonian Eq.~\eqref{trans1} on varying energy scales;  
		data shown is for a coupled transmon chain of length $N=10$.
		Panel (a) illustrates the clustering of levels into energy bundles corresponding to the total number of bosonic excitations.
		Panel (b) zooms in to the 5-excitation band, which upon further enlargement in panel (c) reveals level repulsions that
		become particularly visible for larger couplings. 
		In panel (c) we also mark in red a number of computational states (identified in this energy window at vanishing coupling $T=0$).
		The further zoom-in of panel (d) traces one such computational state through a sequence of avoided level crossings.
		}
	\label{NSF}
 \end{figure*}

This model has been very concretely realized in experiments in many labs in recent
years, but notably also in the many chips that have been made available for use in
the IBM cloud service \cite{IBMQuantumCloud}. These devices of the ``Falcon" and
``Hummingbird" generations have employed transmons laid out in the heavy hexagon
lattice geometry of Fig.~\ref{Fig:real-IBM-chips}(a). While these devices have fixed
values of the coupling parameter $T$ and of the charging energy $E_C$, their
Josephson energy $E_J$ varies from transmon to transmon. 
This effectively random variation is, in fact, crucially required to prevent the buildup of inter-transmon resonances, and the  compromising of quantum information; its role in the physics of present-day transmon device structures, with  insights drawn from  many body localization theory, is the central theme of this paper.

Before addressing the physics of the full model Eq.~(\ref{trans1}), let us consider its  low energy limit.
Applying a sequence of approximations  
(series expansion of the Josephson characteristic, rotating wave approximation) one arrives at the effective Hamiltonian

\begin{align}
H&= \sum_i \nu^{\mathstrut}_i \,a_{i}^{\dagger}a^{\mathstrut}_{i} -\frac{E_C}{2}\sum_i a_{i}^{\dagger}a^{\mathstrut}_{i}(a_{i}^{\dagger}a^{\mathstrut}_{i} + 1) \nonumber\\
&\qquad + \sum_{\langle i,j\rangle} t^{\mathstrut}_{ij} (a^{\mathstrut}_{i}a_{j}^{\dagger}+a_{i}^{\dagger}a^{\mathstrut}_{j}),\cr
&\nu_i \equiv \sqrt{8E_{J_i}E_C}, \qquad t_{ij}=\tfrac{T}{4\sqrt{2 E_C}}\sqrt[4]{E_{J_i}E_{J_{j}}} \,.
\label{BH}
\end{align}
To leading order, this model describes the transmon as a harmonic oscillator, where the above choices of energy scales place the frequencies $\bar\nu_i\approx 6$~GHz on average in the middle of a microwave frequency band, convenient for  precision control. 
The attraction term, a remnant of the $\cos$-nonlinearity, is considerably smaller than the average harmonic term, which is desired for transmon operation. Finally, the characteristic strength of the nearest neighbor hopping coefficients, $ |t_{ij}|\approx  \tfrac{T}{4\sqrt 2}\sqrt{\frac{E_J}{E_C}}$  (often called $J$ in the literature), continues to be the smallest energy scale in the problem.

Unless novel  engineering techniques are used (see below),
the above mentioned variations of the Josephson energy, $\delta E_J$, are in the few percent range; thus, at a minimum, there is variation
in oscillator frequencies  $\nu_i$ of around $\delta\nu_i\approx(\bar\nu/2E_J)\delta E_J\approx 120$~MHz, when the typical $E_J\approx12.5$~GHz. This scale is much larger  than the
particle hopping strength, which for the same parameter set is about $|t_{ij}|\approx 6$~MHz.

From the perspective of many body physics, these  variations make Eq.~\eqref{BH}  a reference model for bosonic MBL. For the above characteristic ratio $\delta \nu/|t|\sim 20$
we can hope that the system is in the MBL phase, and we confirm this below. From the perspective of transmon engineering, the frequency spread blocks the buildup of local nearest neighbor or next nearest neighbor inter-qubit resonances. Below, we will discuss how these two perspectives go together  (and where they may depart from each other). However, before turning to the observable consequences of frequency spread, we note that there exist two broad design philosophies for its realization in transmon array structures, schemes {\sc a} and {\sc b} throughout. 

Generally speaking, scheme {\sc a} aims to suppress the frequency spread to the lowest  possible values required for the stability of the structure, or dictated by limits in precision engineering. For example
Fig.~\ref{Fig:real-IBM-chips}(b) shows that the spread of Josephson energies in IBM devices is 
consistent with a Gaussian distribution (with no stringent correlations from site to
site). 
These observations hold true for all current devices whose parameters are documented publicly by IBM \cite{[][{ All calibration data was downloaded on 11/23/2021, except for the data for the recently retired processors `Manhattan' (11/16/2021) and `Dublin' (11/08/2021).}]IBMQuantumCloud}. Figure~\ref{Fig:real-IBM-chips}(c) shows that the variance $\delta E_J$ has in fact remained very constant over 9 realizations of ``Falcon" chips (27 qubits) and the two latest ``Hummingbird" chips (65 qubits). 
This `natural disorder' regime was in use in many generations of quantum
computer processors \cite{8936946} that IBM has provided on its cloud service since
2016.
However, a significant reduction of  disorder has been reported in a very recent line of research at IBM employing high precision laser-annealing
\cite{hertzberg2020laserannealing} as a pattern engineering approach \cite{zhang2020highfidelity}  to be discussed towards the end of the manuscript.

The complementary scheme {\sc b} embraces frequency disorder as a potent means of protecting qubit information, 
and in fact works to effectively \emph{enhance} it. 
Examples in this category include the recent reports from TU Delft on
their extensible module for surface-code implementation~\cite{PhysRevApplied.8.034021}. 
Google devices such as its  53-qubit processor \cite{GoogleQuantumSupremacy} contain engineered frequency patterns whose aperiodic variation effectively realize a form of synthetic disorder, where in addition the qubit coupling $t_{ij}$ is lowered during idle periods.
In this way the ratio $\delta \nu/t$ --- the relevant scale for localization properties --- is drastically enhanced. 

Below, we will consider both scheme {\sc a} and {\sc b}, and investigate the incipient quantum localization and quantum  chaos  present
in the two settings.
In our model calculations, we represent the disorder by drawing
independent samples from a Gaussian distribution with standard deviation $\delta
E_J$, added to the mean Josephson energy $E_J$. 
As a representative {\sc a} case we take
$\delta E_J\approx500$~MHz
(for a Josephson energy of $E_J =12.5$~GHz, giving $\delta\nu\approx 120$~MHz, as above), while for a {\sc b} case we will take
$\delta E_J$ and $\delta\nu$ some 10 times larger (precise numbers are given below). Note that
from this point onward, we continue with the full model Eq.~\eqref{trans1}.

Specifically, for scheme-{\sc a} parameter ranges, the energy eigenvalues of Eq.~\eqref{trans1} cluster into
energy bundles corresponding to the total number of bosonic excitations, as  seen in
Fig.~\ref{NSF}(a). Looking inside the 5-excitation band, we see (in
Fig.~\ref{NSF}(b)) a dense tangle of energy levels. 
However, only some of these levels are
used to perform quantum computations in
quantum processors; the identification of
these levels, as shown in Fig.~\ref{NSF}(c) and discussed in detail below, 
can only be done unambiguously if we are far away from the chaotic phase.

Having QC applications in mind, we are primarily interested in  signatures of quantum
chaos in the `computational subspace' of the bosonic Hilbert space, i.e.\ the space
comprising local occupation numbers $a_i^\dagger a_i^{\mathstrut}=0,1$, defining the  $p$-qubit states for QC. In that Hilbert space sector, the problem reduces to a disordered spin-$\text{\sfrac{1}{2}}$ chain, another paradigm
of MBL. Recent results from the MBL community  indicate that the separation into a
chaotic ergodic and an integrable localized phase is not as straightforward as
previously thought, and that wave functions show remnants of extendedness and
fractality even in the `localized' phase \cite{PhysRevLett.123.180601}.


{\bf Diagnostics. }
In the following, we analyze the Hamiltonian Eq.~\eqref{trans1} with a  combination of different numerical methods   tailored to the description of localized phases: 
\begin{itemize}
    \item \emph{Spectral statistics:} According to standard wisdom, many-body spectra have Wigner-Dyson statistics in the phase of strongly correlated
chaotic states, and 
Poisson statistics  in that of uncorrelated localized states \cite{Serbyn2016}.
Real systems show more varied
behavior,  quantified below  in terms of a Kullback-Leibler divergence. This produces  a charting of  parameter space 
indicating the chaos/MBL boundary and the rapidity  with which the boundary is
approached.
\item \emph{Wave function statistics:} Focusing on the localization regime, we analyze how strongly the eigenstates differ from the localized states of the strictly decoupled system. 
\item \emph{Walsh transform:} We  quantify the correlations between $l$-qubits (known in the QC community as ZZ couplings, and in the MBL community as $\tau$-Hamiltonian tensor coefficients)  by application of a
Walsh transform filter. To the best of our knowledge, this particularly sensitive
tool has not been applied so far to the diagnostics of MBL.
\end{itemize}

We consider a system of $N$ 
coupled transmons in a one-dimensional chain geometry -- a minimalistic setting 
that allows us to probe essential aspects of localization physics and quantum chaos using the above diagnostics and whose computational 
feasibility allows us to map out the broader vicinity of experimentally relevant parameter regimes.
Typical system sizes vary between $N = 5$ and 10 sites, as detailed below.


\begin{figure}[t] 
   \centering 
    \includegraphics{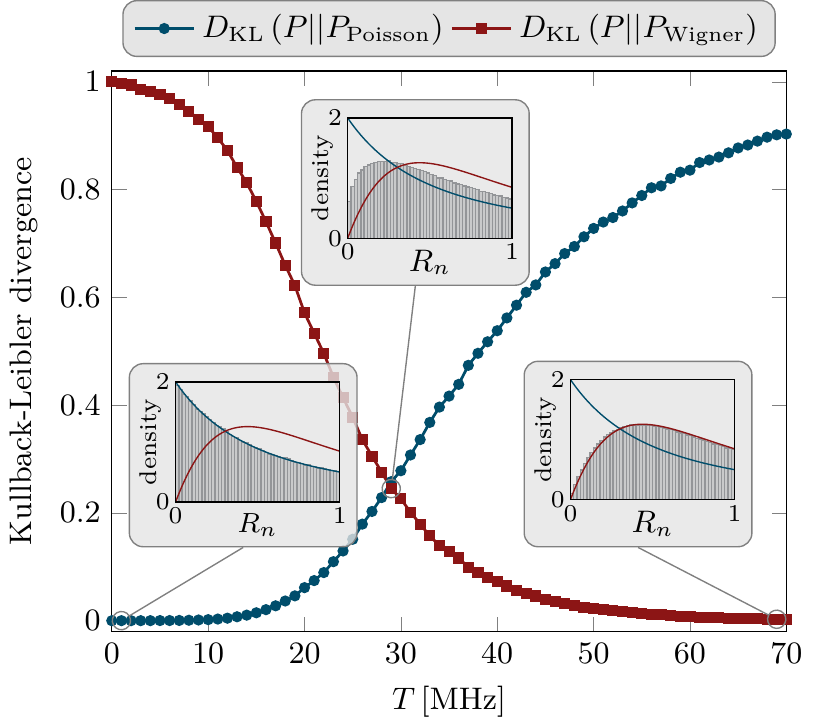}
    \caption{\textbf{Spectral statistics} of a chain of $N=10$ transmons versus the coupling parameter $T$ 
	for fixed average Josephson energy $E_J=44$~GHz and scheme-{\textsc a} disorder ($\delta E_J=1.17~$GHz). 
	These statistics indicate a transition from Poisson statistics (blue) in the MBL
	regime (at low coupling) to Wigner-Dyson statistics (red) in a many-body
        delocalized regime (at large coupling). Shown are normalized Kullback-Leibler
        (KL) divergences $D_{\rm KL}$ (see Eq.~\eqref{eq:KL} in Methods) calculated for the distribution of ratios of consecutive level
        spacings $R_n$ in the energy spectrum, such as the ones illustrated in the insets
        for three characteristic couplings. 
        The KL divergences are normalized such that $D_{\text{KL}} \left( P_{\text{Wigner}}  || P_{\text{Poisson}}  \right)\!=\!1$ and vice versa.
        All results are averaged over at least 2500 disorder realizations.
    }
\label{fig:KL-1Dcut}
\end{figure}

{\bf Spectral statistics. }
We probe the spectral signatures of this coupled transmon
system in an energy bundle of excited states (see Fig.~\ref{NSF}(b)), which are generated by a total of $N/2
= 5$ bit flips and can be viewed as typical representatives in the computational
subspace \footnote{For the $N=10$ transmon chain at hand, this manifold contains a
total of 2002 different states.}.
Zooming in on this mid-energy spectrum, we plot its spectral statistics in the main panel of Fig.~\ref{fig:KL-1Dcut}: 
The KL divergence vanishes
when calculated with  respect to the Poisson distribution for small transmon couplings,
indicating perfect agreement with  what is expected for an MBL phase. This is also corroborated
by the striking visual match of the distributions in the corresponding inset of
Fig.~\ref{fig:KL-1Dcut}. 
But the KL divergence
is maximal when compared to Wigner Dyson
statistics (red curve in Fig.~\ref{fig:KL-1Dcut}). This picture is inverted for
large transmon couplings $T \approx 70$~MHz, where we find an extremely good match to
Wigner-Dyson statistics -- unambiguous evidence for the emergence of strongly
correlated chaotic states. Probably even more important is the fact that these KL
divergences allow us to quantify proximity to the diametrically opposite regimes
for all intermediate coupling parameters. 
This includes a region of `hybrid statistics' around the crossing point of the two curves,
indicating an equal distance from both limiting cases, which we will discuss in more detail below.

\begin{figure}
    \centering
    \includegraphics{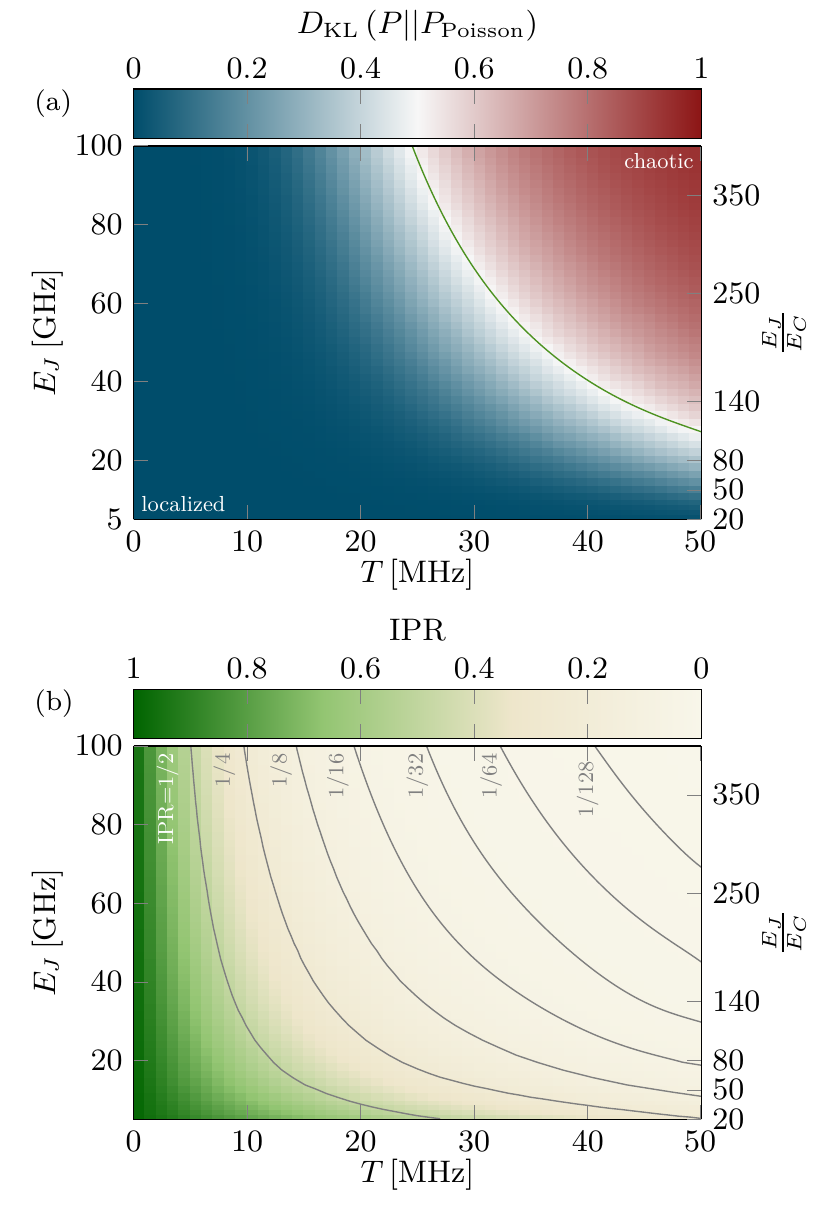}
    \caption{{\bf Phase diagrams} in the plane spanned by the Josephson energy $E_J$ and the transmon coupling  $T$ 
    		for scheme-\textsc{a} parameters.
    		The upper panel summarizes the {\em spectral statistics} by plotting the Kullback-Leibler divergence with  respect to 
		the Poisson distribution, identifying an MBL regime (blue) for small couplings and a quantum chaotic regime (red)
		following Wigner-Dyson statistics for large couplings.
		The lower panel summarizes the {\em wave function statistics} by color-coding the inverse participation ratio (IPR)
		showing a fast drop to values below 1/2 already for moderate coupling strength.
		The grey lines indicate contour lines of constant IPR.  All results are averaged over at least 2000 disorder realizations.  
		The spread of the Josephson energies varies from $\delta E_J \sim 0.4~$GHz for $E_J=5~$GHz to $\delta E_J \sim 1.7~$GHz
		 for $E_J=100~$GHz (see Methods for details).}
    \label{fig:2D-KL-IPR}
\end{figure}

By way of this KL divergence one can then map out an entire phase diagram, e.g.\ in the plane spanned by
varying values of the transmon coupling and Josephson energy, while fixing the charging energy as shown 
in Fig.~\ref{fig:2D-KL-IPR} (for scheme-{\sc a} parameters).
This allows us to clearly distinguish the existence of two regimes, the expected MBL phase (colored in blue) for small 
transmon coupling and a quantum-chaotic regime (colored in red), where the level statistics 
follow Wigner-Dyson behavior (with delocalized, but strongly correlated states) 
for sufficiently strong transmon couplings. It is this latter regime that one surely wants to avoid in any
experimental QC setting. But before we discuss the experimental relevance of our results, we want
to characterize more deeply the quantum states away from this chaotic regime using additional diagnostics.


{\bf Wave function statistics. }
One particularly potent measure of the degree to which a given wave function is localized or delocalized,
is its inverse participation ratio (IPR), i.e.\ the second moment 
\begin{equation}
	{\rm IPR} =  \sum_{\{n\}}  \left\langle|\psi_n|^4\right\rangle \,,
\end{equation}
where the angular brackets denote averaging over disorder realization, and $\sum_{\{n\}}$ is symbolic notation for the summation over a chosen basis (in the present case, the Fock basis).
An IPR of 1 indicates a completely localized state (as in our example for vanishing coupling $T=0$),
while an IPR less than 1 indicates the tendency of a wave function towards delocalization \cite{RevModPhys.80.1355}. 

Here we consider the IPR measured as an average over
all states in one of the energy bundles illustrated in Fig.~\ref{NSF}(a),
e.g.\ the manifold of typical states with
$N/2=5$ bit flips considered in the spectral statistics above. Figure \ref{fig:2D-KL-IPR}(b) shows the IPR in the same parameter space as in (a).
What is most striking here is that the IPR rapidly decays -- the contour lines in the panel indicate exponentially
decaying levels of $1/2, 1/4, 1/8, \ldots 1/128$ -- showing that the wave functions quickly delocalize.
Note in particular, that the IPR has dropped to a value
of less than 10\%  in the region of `hybrid statistics'
identified in the level spectroscopy above.


{\bf Walsh-transform analysis. }
The MBL phase is the right place to be for quantum computing, since computational qubits (the $l$-qubits above) retain their identity there. But, as indicated by the drop of IPR, even the localized phase may be problematic. We therefore apply another diagnostic that is specifically adapted to identifying problems with running a quantum computation in the MBL phase. It begins with the expectation, announced in \cite{PhysRevB.90.174202,PhysRevLett.111.127201}, that the Hamiltonian of the multi-qubit system, in the $l$-qubit basis, can be expressed as 
\begin{eqnarray}
H&=&\sum_ih_i\tau_i^z+\sum_{ij}J_{ij}\tau_i^z\tau_j^z+\sum_{i,j,k}K_{ijk}\tau_i^z\tau_j^z\tau_k^z+\dots \label{tradMBL}\\
&=&\sum_{\bf b} c_{\bf b}Z_1^{b_1}Z_2^{b_2}...Z_N^{b_N}. \label{qiMBL}
\end{eqnarray}
Eq.~\eqref{tradMBL}, the `$\tau$-Hamiltonian' of MBL theory~\cite{PhysRevB.90.174202,PhysRevLett.111.127201}, embodies the observation that a diagonalized Hamiltonian can be written in a basis of diagonal operators $\tau_i^z$, which are the same as the Pauli-Z operators ($Z_i$) in the quantum-information terminology of Eq.~\eqref{qiMBL}.  Here the sum is over $N$-bit strings ${\bf b}=b_1b_2 \ldots b_N$, where each $b_i$ is 0 or 1.  

A system described by the $\tau$-Hamiltonian can be an excellent data carrier for a quantum computer, particularly if the high-weight terms are small. If only the one-body terms in Eq.~\eqref{tradMBL} are non-zero, the system is an ideal quantum memory: In the interaction frame, defined by the non-entangling unitary transformation $U(t)=\exp(i t\sum_ih_i\tau_i^z)$, all quantum states, including entangled ones, remain stationary.
Unfortunately, the expectation of
MBL theory is that the two-body and higher interaction terms are non-zero and grow as the chaotic
phase is approached. 

We have performed a numerical extraction of the parameters of Eqs.~(\ref{tradMBL}-\ref{qiMBL}) for a 5-transmon chain. We find that problematic departures from full localization do indeed occur already at rather small values of the qubit-qubit coupling parameter $T$. This reinforces the message, in a {\em basis-independent} way, of our IPR study. But this extraction must begin with a very non-trivial step, namely the identification of the qubit eigenenergies of the transmon Hamiltonian. Since this Hamiltonian Eq.~\eqref{trans1} is bosonic, it has a much larger Hilbert space than the spin-$\text{\sfrac{1}{2}}$ view embodied in  
Eqs.~(\ref{tradMBL}-\ref{qiMBL}). The qubit states, those with bosonic occupations limited to 0 and 1, are not separated in energy from the others, but are fully intermingled with states of higher occupancy, as illustrated in Fig.~\ref{NSF}(c). It would thus appear that this truncation is rather unnatural --- but it is in fact crucial to the whole quantum computing program with transmons. It is essential to pick out, from all the eigenlevels $E_\alpha$ of the full Hamiltonian Eq.~\eqref{trans1} as shown in Fig.~\ref{NSF}, just the subset of levels $E_{\bf b}$ that can be associated with a bitstring label ${\bf b}$ (cf. Eq.~\eqref{qiMBL}).

\begin{figure}[t]
   \centering
   \includegraphics{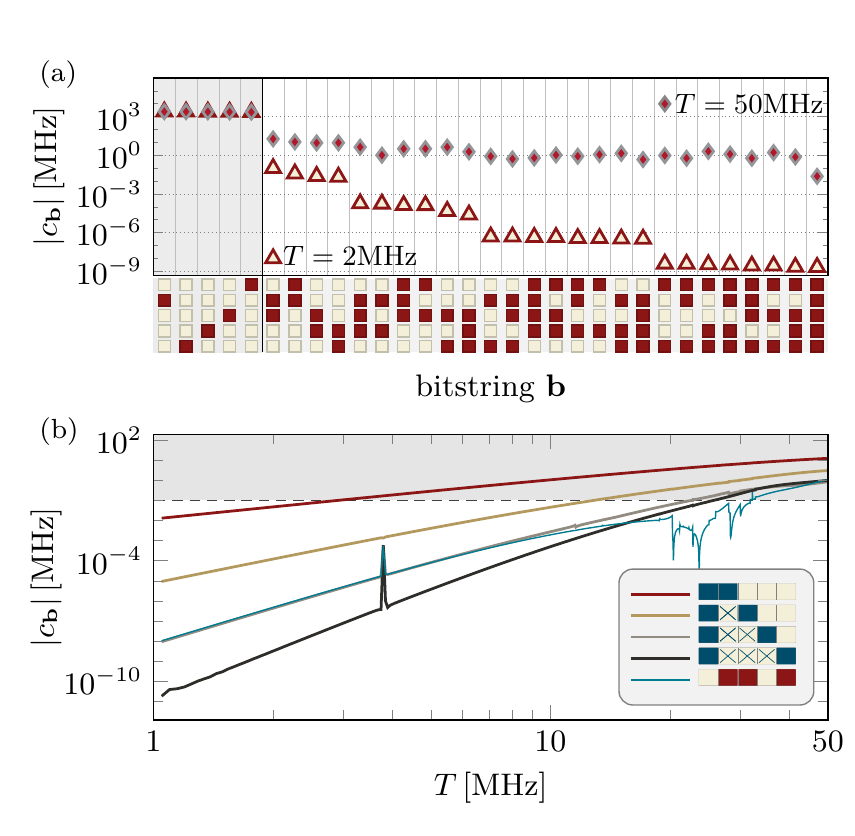}
    \caption{{\bf Walsh-transform analysis.} 
    		(a) Comparison of the $c_{\bf b}$ coefficients of Eq.~\eqref{WH} for a five-qubit chain with scheme-{\sc a} parameters, for two values of the coupling $T$.
		Along the $x$-axis are the 31 different values of the bitstring ${\bf b}$ with at least one non-zero bit.	
		We use a graphical depiction of each bitstring, as a vertical column of five boxes, so that the first bistrings, starting from the left, are 01000, 00001, 00010, etc. With this graphical depiction one can see immediately which of the five transmons are involved in the given $\tau$-Hamiltonian coefficient. The $|c_{\bf b}|$ are sorted from largest to smallest for the $T=2~$MHz data, which reveals a clear hierarchy of strengths according to the maximal distance between two 1's in the bitstring. There is no such systematic behavior for the large-$T$ case (plotted for the same ordering).
		(b) Absolute value of averaged Walsh coefficients as a function of the coupling $T$. The inset introduces a new coloring notation for the bitstrings ${\bf b}$; blue-colored bit boxes indicate that the $c_{\bf b}$ shown is averaged over all cases with the same maximal distance of two 1's. The convention is explained more fully in the Methods section, Fig.~\ref{fig:walsh-manual}. Also shown for comparison is the absolute value of the Walsh coefficient for the specific bitstring ${\bf b}=01101$. The dashed line and the shading above marks the ``danger zone" $|c_{\bf b}|\gtrsim 100$kHz indicated by recent experimental studies on ZZ coupling \cite{PhysRevLett.125.200504}. 
		}
    \label{fig:walsh-groupplot}
\end{figure}

Having performed such a state identification (as discussed in the Methods) and tagged the subset of eigenlevels $E_{\bf b}(T)$ that can be identified as qubit states, the coefficients of the $\tau$-Hamiltonian are easily obtained by a Walsh-Hadamard transform~\cite{Walsh}:
\begin{eqnarray}
c_{\bf b}(T)&=&\frac{1}{2^N}\sum_{\bf b'}
(-1)^{b_1b'_1}(-1)^{b_2b'_2}... (-1)^{b_Nb'_N}E_{\bf b'}(T)\nonumber\\&=&\frac{1}{2^N}\sum_{\bf b'}(-1)^{{\bf b}\cdot{\bf b'}}E_{\bf b'}(T).
\label{WH}
\end{eqnarray}
Being a kind of Fourier transform, the Walsh transform functions to extract correlations, in this case in the correlations of the computational eigenstates (in energy).
Fig.~\ref{fig:walsh-groupplot} shows these coefficients vs. $T$. 
For small $T$, many of the expectations from MBL theory~\cite{PhysRevB.90.174202,PhysRevLett.111.127201} are fulfilled: There is a clear hierarchy according to the locality of the coefficients. Thus, nearest-neighbor ZZ interactions are the largest, followed by second-neighbor ZZ and contiguous ZZZ couplings, and so forth. Jumps occur in these coefficients, initially very small, which arise from the switching of labeling at anticrossings.

The two-body ($J_{ij}$/ZZ) terms are known and carefully analysed in transmon research \cite{PhysRevA.101.052308,PhysRevLett.125.200504,sundaresan2020reducing}. Their troublesome consequences, including dephasing of general qubit states, and failure to commute with quantum gate operations, 
add overhead which is ultimately found to be insupportable, enforcing a 
practical upper limit of $J_{ij}\sim 50$--$100$~kHz. 
This limit, marked (dashed line) in Fig.~\ref{fig:walsh-groupplot}(b), is exceeded already at $T=3$~MHz. Even though the transition to chaos is still a long way off, quantum computing becomes very difficult above this limit.

{\bf Scheme {\sc b}: spread frequency distribution. } 
Other techniques for executing entangling gates leave considerably more freedom to increase the disorder, with $\delta E_J$ values in the GHz range. This option can forestall the growth of problematic precursors of chaotic behavior. A good example of a quantum computer that uses this freedom is the surface-7 device of TU Delft~\cite{PhysRevApplied.8.034021}. During gate operation the qubit frequencies are temporarily tuned into resonant conditions that `turn on' entanglement generation. This is done in a pattern that does not lead to any extensive delocalization. In the 53-qubit quantum computer of Google~\cite{GoogleQuantumSupremacy}, this tuning is also available, but in its operation  an additional strategy is used: extra hardware is introduced to also make $T$ tunable. Being able to set the effective $T$ to zero (although only in a perturbative sense) of course eliminates the problem of delocalization, and in this latest Google work $\delta E_J$ has been returned to a small value. Google made major changes in its `Hamiltonian strategy' in  recent years \cite{GoogleQuantumSupremacy} that have led them to their recent success.
In the Supplementary Material \cite{Supplement}, we discuss scheme-\textsc{b} parameters (as found in recent Delft chips \cite{PhysRevApplied.8.034021}) quantitatively using our three diagnostics.



\begin{figure}[t]
	\hspace*{-0.4cm}
    	\includegraphics[width=\columnwidth]{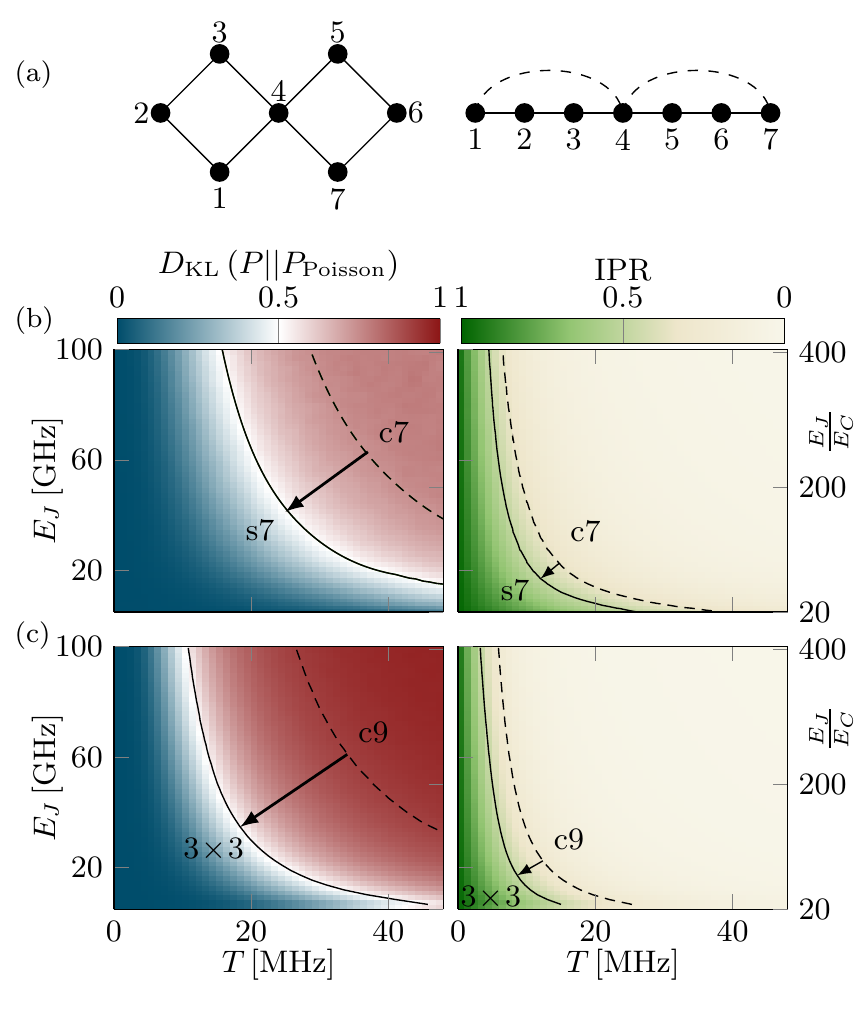}
    	\caption{\textbf{Two-dimensional transmon geometries.} 
    		(b) Phase diagram of seven coupled transmons, coupled in a surface 7 (s7) geometry (indicated on top of the figure).
		The inclusion of two additional couplings in comparison to a chain of seven transmons (c7) leads to significant shifts
		in the phase diagram calculated for {scheme-{\textsc a} } parameters, 
		as illustrated in the left panel for the level statistics and in the right panel for the IPR.
		Shown on the left is the shift of the line indicating where the normalized KL divergence with regard to Poisson
		statistics has increased to 0.5 (see also Figs.~\ref{fig:KL-1Dcut} and \ref{fig:2D-KL-IPR}). 
		On the right we indicate the shift of the line indicating where the IPR drops below 0.5, akin to the lower panel in Fig.~\ref{fig:2D-KL-IPR}.  All results are averaged over at least 1500 disorder realizations. (c) Phase diagram of a $3 \times 3$ transmon array. All results are averaged over at least 2500 disorder realizations. For both geometries, the same scheme-{\textsc a} parameters as in Fig. \ref{fig:2D-KL-IPR} were used.
}
	\label{fig:s7}
\end{figure}

{\bf Transmon arrays in higher dimensions. }
All the conclusions of the last few sections have been reached by calculations for transmons coupled in a one-dimensional chain geometry. However, actual quantum information architectures are two dimensional, and we have therefore also simulated the surface-7 layout~\cite{PhysRevApplied.8.034021,Wallraff_s7_2020} as well as a $3\times 3$ transmon array as minimal examples in this category. The surface-7 chip comprises  a pair of square plaquettes, which is obtained from a chain of seven transmons by including two additional couplings, see top panel of Fig.~\ref{fig:s7}.
(Google's 53-qubit layout extends this principle to a large square array extension.)

The study of two dimensional geometries, or of one dimensional arrays shunted by long range connectors, is motivated by the realization that the 
case of strictly one dimensional chains is exceptional: in one dimension, `rare fluctuations' with anomalously strong local disorder amplitudes may block the correlation between different parts of the system, enhancing the tendency to form a many-body localized state. In higher dimensions, such roadblocks become circumventable, which makes disorder far less efficient in inhibiting quantum transport. For an in-depth discussion of the effects of dimensionality on MBL we refer to Ref.~\cite{PhysRevX.5.031033}. 

Our simulations of the surface-7 and $3 \times 3$ architectures, where we have chosen scheme-{\sc a} parameters, are summarized in Fig.~\ref{fig:s7}~(b) and (c). The spectral and wave function statistics data 
indicate that, not surprisingly, chaotic traces are rather more prominent than in the one-dimensional simulations
(and despite the fact that to get the surface-7 geometry, we nominally added only two extra couplings in comparison to a purely one-dimensional geometry, see Fig.~\ref{fig:s7}(a)). 
The bottom line is that the comfort zone introduced by disorder schemes is considerably diminished when
including higher-dimensional couplings.



\begin{figure} 
    \includegraphics[scale = 1]{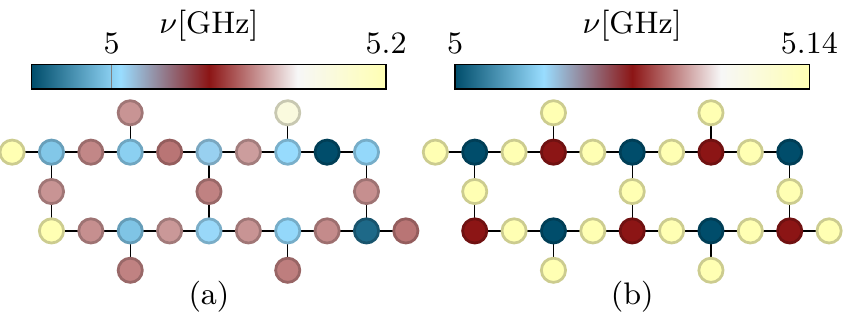}
    \caption{{\bf Frequency alternation patterns} for scheme-{\sc a} architectures on the heavy-hexagon geometry \cite{chamberland2020, hertzberg2020laserannealing}: (a) A-B-A-B pattern. (b) A-B-A-C pattern. Note that while (b) is an idealized structure, the A-B pattern of (a) is one that is currently implemented in experiment \cite{zhang2020highfidelity}, and thus with slighlty imperfect setting of frequencies, visible especially on the far left and right of the device.
    }
    \label{ABABCm}
\end{figure}

{\bf Qubit frequency engineering. }
Until very recently, process variations in $E_J$ have led to an inevitable spread
in qubit frequencies, as described by the effectively Gaussian distributions employed
above (see Fig.~\ref{Fig:real-IBM-chips}(b)). However, the development of a high precision laser-annealing
technique \cite{hertzberg2020laserannealing}  (LASIQ, see Fig.~\ref{Fig:real-IBM-chips}(c)) has changed the
situation and is opening the prospect to clone qubits  with unprecedented precision.
IBM proposes \cite{zhang2020highfidelity} to use this freedom and realize arrays with
A-B-A-B, or A-B-A-C (Fig.~\ref{ABABCm}) frequency alternation, effectively blocking unwanted
hybridization between neighboring qubits. However, even then a residual amount of random
frequency variation remains essential for the functioning of the device. For example,
a perfect A-B-A-B sequence would block nearest neighbor hybridization at the expense
of creating dangerous resonance between \emph{next nearest} qubits; more formally, perfectly  translationally invariant arrays would have extended Bloch eigenstates, different from the localized states required for computing. 

The question thus presents itself how to optimally navigate a landscape defined by the extremes of Bloch extended, chaotic, and many body localized wave functions for absent, intermediate, and strong disorder, respectively. In the Supplementary Material \cite{Supplement} we apply the diagnostic framework introduced earlier in 
the paper to address this question in quantitative detail. To summarize the results, we observe that for diminishing disorder the transmon-array Fock space disintegrates into a complex system of mutually decoupled subspaces, reflecting the complexity of the A-B or A-B-A-C ``unit cells''. The  strength of the residual disorder 
determines whether wave functions are chaotically extended or localized within these structures. Employing the inverse participation ratio as a quality indicator, we find that recent IBM engineering succeeded in hitting the optimum of near localized states with IPRs close to unity. However, it is equally evident that further reduction of the disorder would delocalize these states over a large number of qubit states and be detrimental to computing; disorder remains an essential resource, including in devices of highest precision.



{\bf Discussion. }
The subject of this study has been an application of state-of-the-art methodology of many-body localization theory to realistic models of present-day transmon computing platforms. In the mindset of the localization theorist, all transmon quantum information hardware has in common that it operates in a regime where the tendency of quantum states to spread by inter-qubit coupling is blocked by the detuning of qubit frequencies --- the many-body localized phase. Within this phase, there is considerable freedom for the realization of localization-protected architectures; different strategies include {\sc a}: weak coupling at weak detuning (IBM), or {\sc b}: strong, intentionally introduced detuning interspersed by sporadic and incomplete couping (Delft/Google). On this background, we have explored the integrity of different localized phases, in view of the omnipresent phase boundary to a chaotic sea of uncontrollable state fluctuations in the limit of too weak detuning and/or too strong coupling.

The single most important insight of this study is just how extended the twilight
zone of partially compromised quantum states already is before reaching the boundary to
hard quantum chaos. One may object that the existence of a crossover zone is owed
to the smallness of the transmon arrays of 5-10 units studied in this work. However,
it has to be kept in mind that the dimension of the random Hilbert spaces in which
the many body quantum states live is exponential in these numbers and large by any
(numerical) standard of localization theory. This indicates that `finite-size effects' in these systems are notorious and must be kept in mind for computing architectures of technologically relevant scales.

A second unexpected finding is that
early indicators of chaotic fluctuations show in different ways in different
observables. Among these, the least responsive observable is many body spectral
statistics, the most frequently applied  diagnostics of the MBL/chaos
transition. However,
the computational states themselves respond far more
sensitively to departures from the limit of extreme localization. We have observed
this tendency in the standard observable for wave function statistics, the inverse
participation ratios, where Fig.~\ref{fig:2D-KL-IPR} shows tendencies to strong wave
function spreading already in parameter regimes where spectral statistics suggests
complete safety. Surprisingly, however, the Walsh transform diagnostic --- which is
uncommon in MBL theory, but highly relevant as an applied quality indicator for the
integrity of physical qubit states --- responds even more sensitively to parameter
changes away from the deep localization limit. Expressed in the language of quantum information
technology, it indicates strong ZZ coupling and the onset of ZZZ coupling already in
regimes where the participation ratios are asymptomatic.

We note that the most recent experimental work has been strongly focused on the necessity to break the linkage between larger $T$ coupling and the appearance of ZZ coefficients. On the hardware side, ingenious new coupler schemes \cite{kandala2021} show nearest-neighbor ZZ reduced to well below the reference level of 100kHz of earlier work. It is further shown that control techniques, involving advanced refocussing strategies, also strongly diminishes the effect of nearest-neighbor ZZ for a given $T$ \cite{wei2021quantum}. But our work provides a warning that these innovations may ultimately not be enough: all other couplings in the $\tau$-Hamiltonian hierarchy, including next neighbor ZZ and ZZZ, remain neither diagnosed nor ameliorated in the current experiments.

Third and finally, our study of precision engineered qubit arrays has revealed a general structure which we
believe is ubiquitous in weakly disordered transmon arrays: a restructuring of the
`total Hilbert space' into subspaces which  are weakly cross-correlated, but
strongly (`chaotically') correlated  within themselves.  These hierarchies include
spaces of fixed excitation number, or still further refined subspaces of these, distinguished by specific qubit permutation 
symmetries. Where this splintering occurs, a twofold task presents itself: first identify the pattern of relevant spaces, second apply the diagnostic tools discussed in earlier sections within these small spaces. 
Particular care must be exercised in cases where the relevant spaces overlap in energy. Absent considerable inter-space correlations one may then be tricked into the conclusion of Poissonian level statistics (localization!) where in fact wave functions are chaotically extended over the basis of a computationally relevant space. In the Supplementary Material \cite{Supplement} we detail all this intricate phenomenology, using the case study of the LASIQ engineered A-B-A-B pattern as an example illustrating these principles and the development of reliable predictions for the integrity of qubit wave functions.

What is the applied significance of these observations?  One bottom line is that further reduction of the frequency variance may be dangerous.
All transmon based quantum technology operates in a tension field defined by the desire to optimally protect (detune) and efficiently operate (couple). There are different approaches to resolving this conflict of interests, scheme {\sc a} `weak coupling/weak detuning' and scheme {\sc b} `transient-incomplete coupling/strong detuning' defining two master strategies. Our study  indicates that the {\sc a} approach is  more vulnerable to chaotic fluctuations. 
We go so far as to speculate that it may not sustain the generalization to larger and two-dimensionally interconnected array geometries required by more complex applications. 
However, regardless of what hardware is realized, the findings of this work indicate that the shadows of the chaotic phase are much longer than one might have hoped and that careful scrutiny of chaotic influences should
be an integral part of future transmon device engineering.



\acknowledgments
We thank S. B\"orner for collaboration on an initial project \cite{borner_simon-dominik_classical_2020} 
studying incipient chaos in classical transmon systems.
We acknowledge partial support from the Deutsche Forschungsgemeinschaft (DFG) 
under Germany's Excellence Strategy Cluster of Excellence Matter and Light for Quantum Computing (ML4Q) EXC 2004/1 390534769
and within the CRC network TR 183 (project grant 277101999) as part of projects A04 and C05. 
The numerical simulations were performed on the CHEOPS cluster at RRZK Cologne and 
the JUWELS cluster at the Forschungszentrum J\"ulich.




\section*{Methods}

{\bf Spectral statistics via Kullback-Leibler divergence. }
To quantitatively analyze the spectral statistics, 
we look at the distribution of the ratios of adjacent level spacings $r_n = \Delta E_n / \Delta
E_{n+1}$ \cite{PhysRevB.75.155111} (in order to avoid level unfolding) by comparing to what is expected for these ratios
in Poisson or Wigner-Dyson statistics. This is often done via qualitative
observations, such as focusing on the limit of $r \to 0$, where the distribution
exhibits a maximum for Poisson statistics, but  vanishes for Wigner-Dyson
statistics (see, e.g., the insets of Fig.~\ref{fig:KL-1Dcut}). 
However, a recent study~\cite{PhysRevResearch.3.013023}  
(of a Fock space localization transition)
has shown that such inspections may trick one into false conclusions and that the Kullback-Leibler (KL) divergence \cite{mezard_information} provides a far more reliable quantitative alternative. The KL divergence  
\begin{equation}
	D_{\rm KL}(P||Q) = \sum_k p_k \log\left(\frac{p_k}{q_k}\right) \,,
	\label{eq:KL}
\end{equation}
defines  an entropic measure quantifying the logarithmic difference between two
distributions $P$ and $Q$. In our case, the $p_k$ are extracted from the numerical
spectrum for a given set of parameters, while the $q_k$ follow one of the two
principal spectral statistics considered here.
Note that in Fig.~\ref{fig:KL-1Dcut} we plot $R_n = \min\left( r_n, 1/r_n \right)$ in order to restrict to the range $[0,1]$.
 
\begin{figure}[t]
	\centering
	\includegraphics{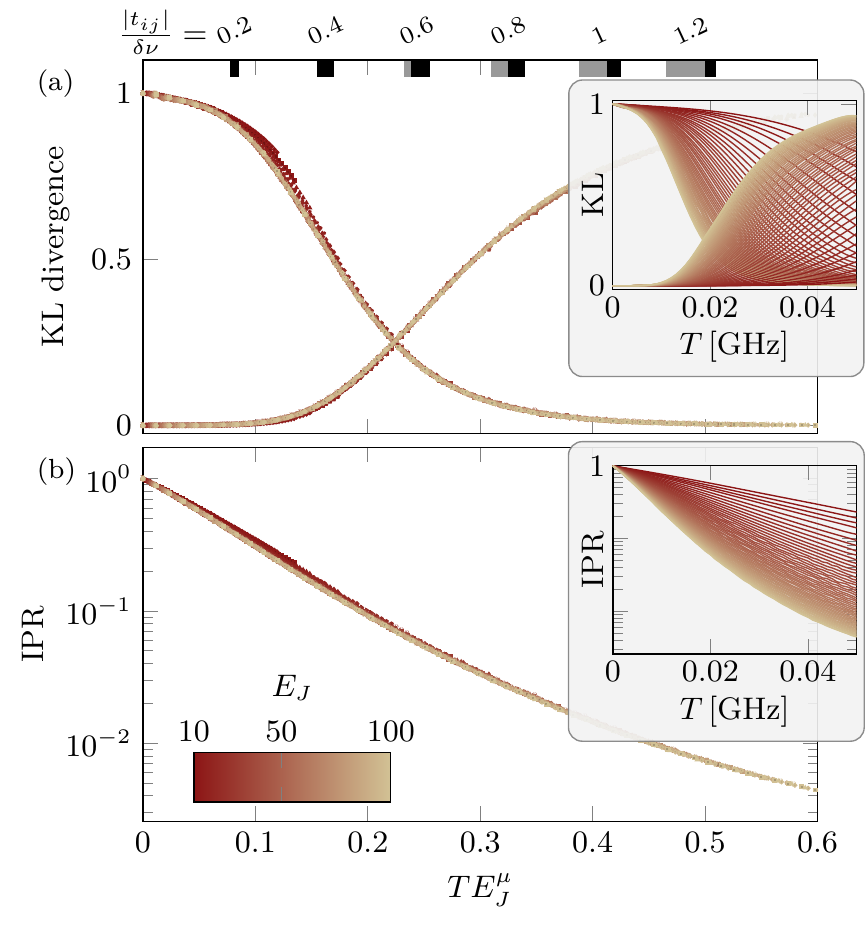}
    \caption{\textbf{Data collapse} of the individual traces of the KL divergence and IPR, 
    		  underlying the phase diagrams of Fig.~\ref{fig:2D-KL-IPR} and shown in the two insets,
		  by rescaling the coupling parameter with  respect to the Josephson energy as 
		  $T \to T  E_J^\mu$ with an exponent $\mu \approx 0.54$. 
		  As $T E_J^\nu \sim |t|/\delta \nu \cdot E_J^{\mu - 1/2}$, a $T E_J^\mu$ interval, whose lower (upper) bound is determined by the minimal (maximal) $E_J$ value, belongs to each $|t|/\delta \nu$ value. The areas shaded in black indicate the $T E_J^\mu$ range associated with the $|t|/\delta \nu$ label next to it. The upper boundary corresponds to $E_J = 100$ GHz. The lower boundary is obtained for the smallest $E_J$ for which data points at that particular $|t|/\delta \nu$ exist. The lower boundary of the gray-shaded interval corresponds to $E_J=10$ GHz, the smallest Josephson energy considered in the data collapse.}
    \label{fig:DataCollapse}
\end{figure}

{\bf Data collapse and phase transition. }
A particularly consistent picture emerges if one performs a simple rescaling of the
numerical data in both panels of Fig.~\ref{fig:2D-KL-IPR}. As shown in
Fig.~\ref{fig:DataCollapse}, the individual traces of both the KL divergence
and the IPR for varying values of the Josephson energy $E_J$ (shown in the insets)
all collapse onto one another when rescaling the coupling parameter $T \to T  E_J^\mu$
with the exponent $\mu$ being the single free parameter. Such a data collapse is
typically considered strong evidence for the existence of a {\em phase transition},
i.e.\ we can manifestly separate the MBL phase for small transmon couplings from a
truly chaotic phase for sufficiently large couplings. 

This also allows us to mark a
Rubiconian line  on our phase diagram (indicated by the green line in the top
panel of Fig. \ref{fig:2D-KL-IPR})
that should not be crossed in any
quantum computation scheme, as all exquisitely prepared quantum information would be
 lost quickly upon entering the realm of quantum chaos lying beyond.
The data collapse at a value $\mu\simeq 0.5$ follows from a simple argument:
thinking of the wave functions as states  living on a high dimensional lattice
defined by the occupation number configurations $n=(n_1,n_2,\dots,n_N)$ ($n_i=0,1$
for the computational subspace),  individual occupation number states $n$ are connected to a large
number  of neighbors via the `hopping matrix elements', $t_{ij}$. Wave functions
hybridize over a pair of configurations $n,m$ , provided $|t_{ij}|\gtrsim |\Delta \epsilon_{nm}|$,
where $\Delta \epsilon_{nm}$ is  the energy difference between the two  configurations in the
limit $t\to 0$. Inspection of Eq.~\eqref{BH} shows that $\Delta \epsilon_{nm} $ depends on the $E_J$s as
$E_{J_i}^{1/2}-E_{J_j}^{1/2}\sim E_J^{-1/2} (E_{J_i}-E_{J_j})$. In  the analysis of Fig.~\ref{fig:DataCollapse}, the
random deviations are scaled such that $(E_{J_i}-E_{J_j})\sim E_{J}^{1/2}$, such that
$\Delta \epsilon_{nm}$ is effectively independent of $E_J$. However, $t_{ij}\sim T
E_J^{1/2}$, indicating that $T E_J^{1/2}$ is the relevant scaling variable for the
transition where the two parameters $T$ and $E_J$ are concerned.

{\bf State identification for Walsh transform. }
We find that there is a workable procedure for identifying computational states in the bosonic spectrum, which however starts
to become problematic long before we reach the MBL-chaotic phase boundary. We adopt the
following assignment procedure: at $T=0$ all states have exact bosonic quantum
numbers, so the $2^N$ eigenstates with bitstring label $\bf{b}$ (the Walsh transforms
of the bitstrings in Eq.~\eqref{qiMBL}) are immediately identified there. 
We increase
$T$; as long as no near-crossings of energy levels occur, the labeling remains
unchanged. We then find that the first near-crossings that occur have the character
of  isolated anti-crossings with very small gaps. In this situation we can
confidently associate the label ${\bf b}$ with the diabatic state (i.e., the one that
goes straight through the anticrossing). We show this by the coloring of Fig.~\ref{NSF}(c). Empirically, this identification procedure works through many
anticrossings, up to about half way to the phase transition; gradually, gaps become
larger, and identification of qubit states becomes more ambiguous (Fig.~\ref{NSF}(d)). Naturally, in the
chaotic phase, eigenstate thermalization says that there is no hope of consistently
identifying any eigenstates as information-carrying multi-qubit states.

\begin{figure}[tbh]
    \centering
	\includegraphics{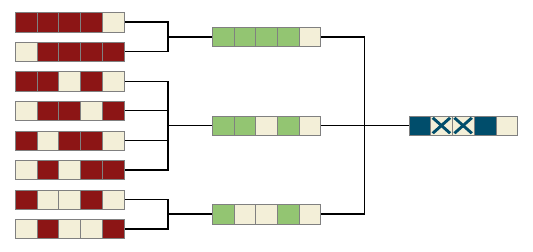}
    \caption{\textbf{Color-coding of Walsh coefficients} for a system of five transmon qubits.
     		The first column shows the individual qubit assignments (light color = `0', red color = `1'), 
		the second and third columns indicate ways to average coefficients  
		according to the enclosing brackets.}
    \label{fig:walsh-manual}
\end{figure}

For clarity of visualization, we do not show all Walsh coefficients $c_\textbf{b}$ in Fig.~\ref{fig:walsh-groupplot}, and in Fig.~\ref{fig:walsh-google} of the Supplementary Material \cite{Supplement}, but average over those with bitstring labels of equal maximal distance of two 1's, as illustrated in Fig. \ref{fig:walsh-manual} for a five transmon chain with the coefficents of maximal distance four.

{\bf Simulation parameters. }
All simulation data shown in the main manuscript were calculated for $E_C = 0.25$~GHz. For scheme-{\textsc a} disorder, we use $\delta \nu \sim E_C / 2$, resulting in a Josephson energy spread $\delta E_J = \sqrt{E_CE_J / 8}$. The Walsh-transform analysis was performed for $E_J=12.5$~GHz.


\clearpage

\section{Supplementary Material}

{\bf Experimental frequency disorder. } In the main manuscript we consider two
principle approaches to the inclusion of frequency disorder in the design of transmon
array devices, schemes {\sc a} and {\sc b} above. The fundamental difference between
these two approaches is reflected in the dimensionless parameter $\delta \nu / t$,
i.e. the strength of frequency disorder relative to the bare transmon coupling, which
we have discussed as a proxy for the stability of many-body localization physics. Here
we want to put the experimental approaches of Google and IBM into the context of these
model classifications.
  
Let us start by summarizing typical parameters for IBM's cloud devices \cite{IBMQuantumCloud}
which have been guiding us in the discussion of the main manuscript
\begin{itemize}
	\item $\delta \nu = 70$--$130$~MHz
	\item $t \approx 3~$MHz 
	\item $ \delta \nu / t \approx $ {30}
	\item gate time: usually $\sim400$ ns, $100$–$200~$ns are possible \cite{PhysRevA.93.060302}.
\end{itemize}
This design scheme is, at its core, geared towards {\em minimizing disorder} and has been dubbed
scheme {\sc a} in the main manuscript.
  
Before turning to its current generation of tunable coupler devices, Google's 
quantum devices \cite{Barends2014}  (as well as recent Delft chips~\cite{PhysRevApplied.8.034021}) operated at the order of  
\begin{itemize} 
    \item $\delta \nu \approx 1 \mathrm{GHz}$
    \item $t = 30~$MHz  
    \item $ \delta \nu / t \approx $ {30}
    \item gate time: 40 ns
 \end{itemize} 
While tolerating a much stronger frequency spread (which brings this setting closer
to design scheme {\sc b}), the  magnitude of the dimensionless parameter  $\delta \nu
/ t \approx 30$ is close to what is seen in the IBM design above. An important
distinction, however, is that the variability of the equally enhanced transmon
coupling $t$ is much more restricted, as it has to remain well below the absolute value
of the typical charging energy $E_C = 250$~MHz. On the other hand, its already large
value of  $t = 30$~MHz leads to notably shorter gate operation times as in the IBM
case.

\begin{figure}
    \includegraphics[scale=1]{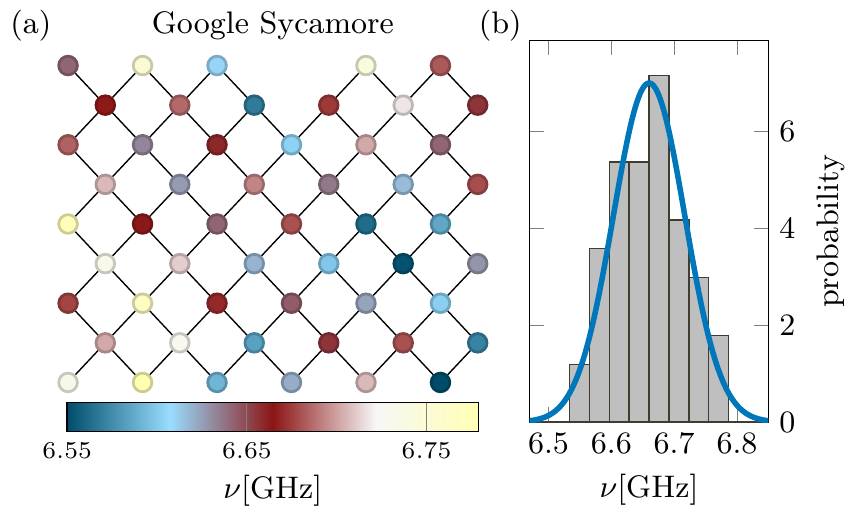}
    \caption{{\bf Experimental parameters of Google's  transmon array.} 
    	(a) Layout of the 53-qubit transmon array ``Sycamore".
    	  	The coloring of the qubits indicates the variation of frequencies 
	 	 which is largely uncorrelated in space. 
   	 (b) Spread of the frequencies 
	 	plotted for the ``Sycamore" chip, consistent with a Gaussian distribution (solid line).
	 }
    \label{fig:sycamore}
\end{figure}

These settings should be contrasted to the current generation of Google devices, such
as the ``Sycamore" processor (whose frequency disorder is visualized in
Fig.~\ref{fig:sycamore}). The  introduction of tunable couplers
\cite{PhysRevApplied.10.054062}  allowed Google to
increase the relative strength of disorder by more than an order of magnitude.
The device settings are \cite{GoogleQuantumSupremacy,Neill2018}
\begin{itemize}
	\item $\delta \nu \approx 60$~MHz
	\item $t <  0.05$~MHz
	\item $\delta \nu / t >$ {1200}
	\item gate time: 12 ns
\end{itemize}
This is a clear-cut example of design scheme  {\sc b}; a significant amount of disorder
protecting  the integrity of information via the principles of MBL.

\begin{figure}[h!]
    \centering
   \includegraphics[scale=0.9]{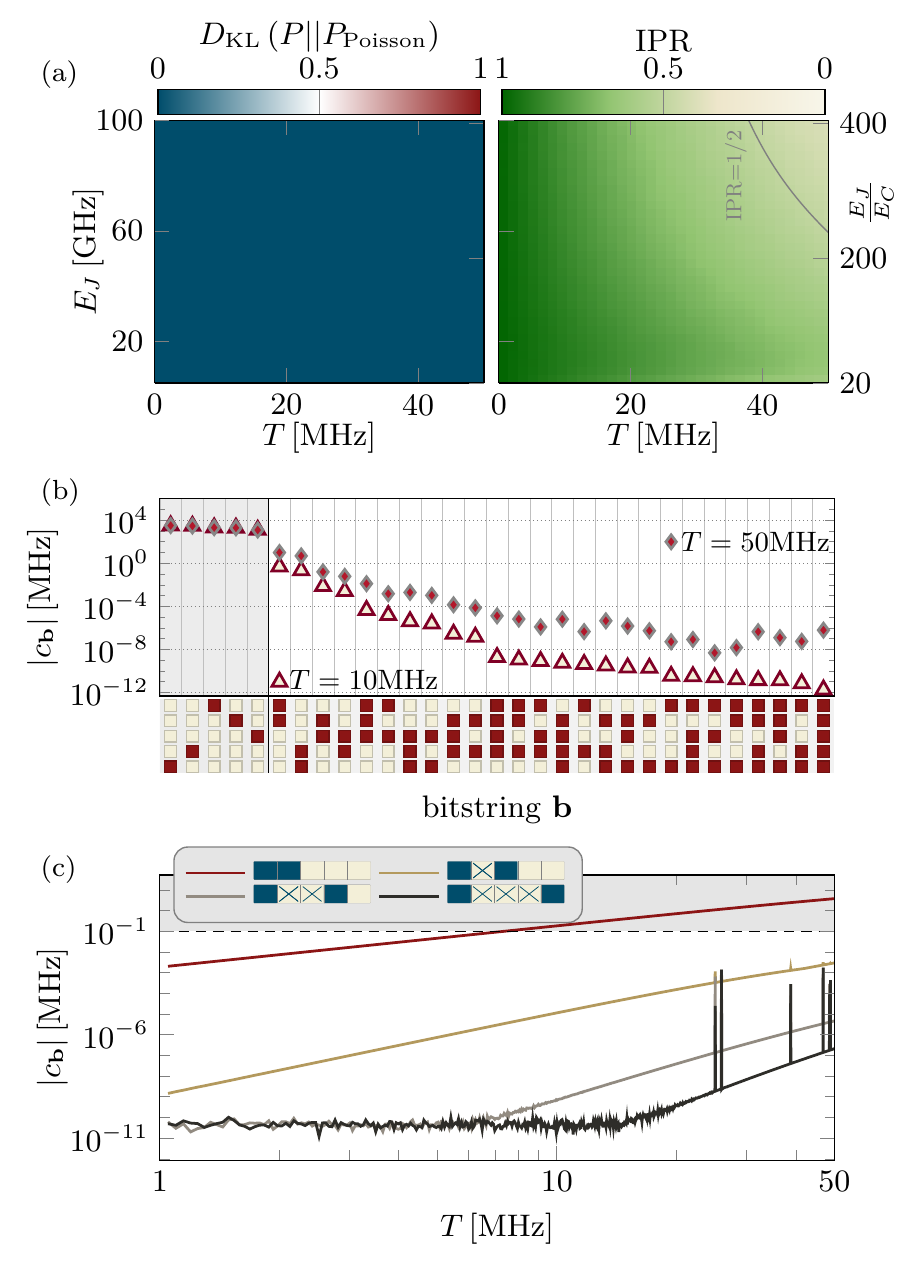}
    \caption{{\bf Summary of diagnostics for scheme-{\textsc b} parameters. }
    		(a) Phase diagram in terms of spectral statistics (left) and IPR (right), akin to Fig.~\ref{fig:2D-KL-IPR} in the main manuscript.
		(b) and (c) Walsh analysis, akin to Fig.~\ref{fig:walsh-groupplot} in the main manuscript.  
		The results in (a) are averaged over at least 2000 disorder realizations. We use $E_C = 250~$MHz and $\delta \nu \sim 6 E_C$, corresponding to $\delta E_J = \sqrt{18 E_C E_J}$. The spread of the Josephson energies thus varies from $\delta E_J \sim 3.4~$GHz for $E_J=5~$GHz ($E_J$ is bounded from below to ensure $E_J/E_C>20$) to $\delta E_J \sim 15~$GHz for $E_J=100~$GHz.
The Walsh-transform analysis in (b) and (c) was performed for $E_J=12.5$~GHz.}
    \label{fig:walsh-google}
\end{figure}

{\bf Scheme {\sc b} diagnostics. }
To complement our analysis for IBM's experimental parameters in the main text, let us consider parameters 
corresponding to scheme {\sc b}, exemplified in 
recent Delft chips \cite{PhysRevApplied.8.034021}. We summarize the three diagnostics of our analysis in Fig.~\ref{fig:walsh-google}. The phase diagrams of Fig.~\ref{fig:walsh-google}(a) show that the MBL-chaos transition has retracted to much larger $T$, with the KL calculation showing no significant departure from Poisson behavior. A minor drop in the IPR indicates dressing effects  much smaller than in the  scheme-\textsc{a} case. However, the Walsh diagnostic for a disorder realization with $\delta E_J =7.5$~GHz, summarized in Fig.~\ref{fig:walsh-google}(b) and (c), shows that trouble is still around the corner: Higher-order terms of the $\tau$-Hamiltonian are still present (albeit at a relatively lower level) and the  ZZ danger threshold (dashed line) remains in sight. These values are observed for  $T=9$~MHz, about three times larger than for the natural-disorder parameters (scheme {\sc a}) discussed in the main text. 
(Note that for the experimental transmon coupling of $t \approx  30$~MHz (corresponding to $T \approx 20~$MHz)  the Walsh coefficient $c_{11}$ (indicating the strength of the ZZ coupling) reaches a value 
in rather close agreement to the reported effective qubit-qubit coupling of $J_\text{eff} = 0.3$ MHz at the idle points in experiment \cite{Barends2014}.)

{\bf Qubit frequency engineering.} In its pursuit to engineer  cleaner devices, IBM has  recently introduced a  two-step  design \cite{hertzberg2020laserannealing} wherein 
 the Josephson energy $E_J$ (qubit frequency) is tuned by a laser-annealing  technique ``LASIQ''   after the initial device fabrication process. The primary goal of this approach is to avoid frequency collisions \cite{hertzberg2020laserannealing,PhysRevA.101.052308} (where, e.g., the transition frequencies of nearby qubits become degenerate).

\begin{figure}[h!]
    \includegraphics[scale = 1]{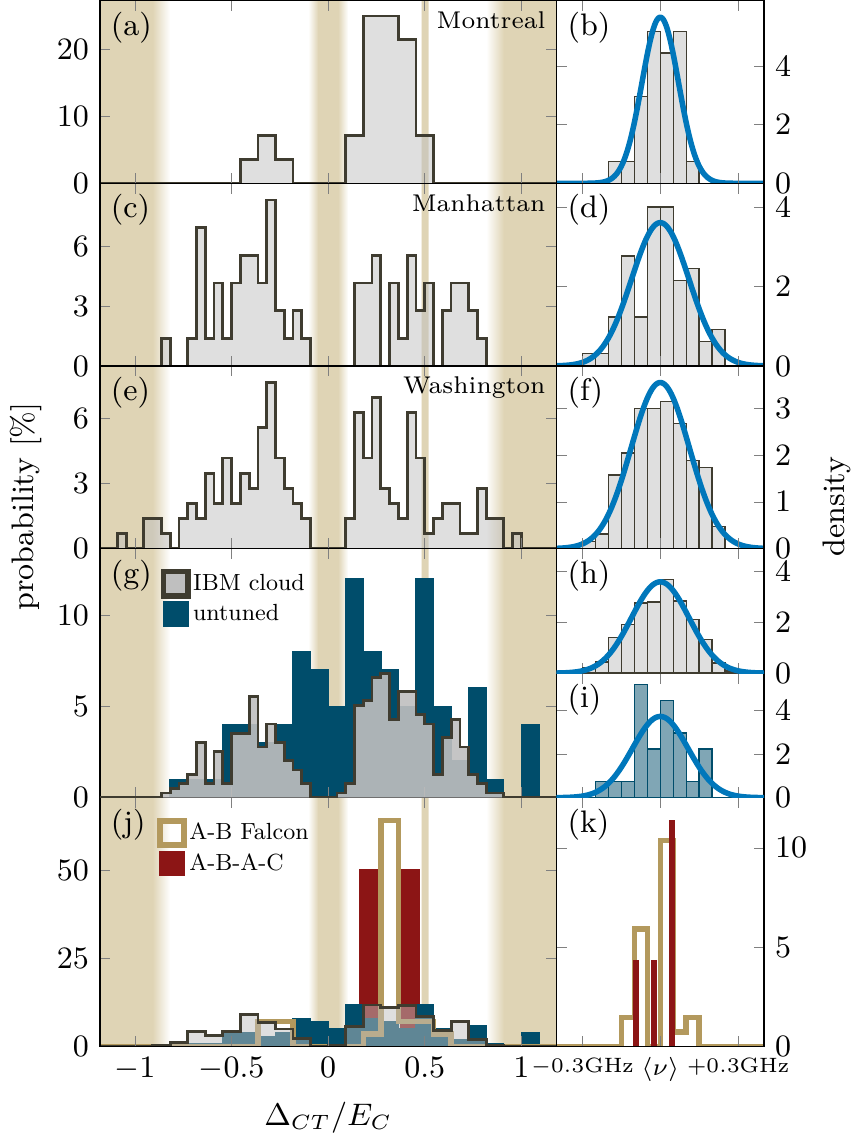}
    \caption{{\bf Overview of IBM cloud devices and LASIQ engineering.}
    	Left column: distribution of NN frequency differences. 
		Right column: distribution of frequencies around their mean values $\langle \nu \rangle$ (histograms) 
		and fitted Gaussian distribution (solid line).
		Tanned areas mark the $\Delta_{CT}$ ranges where NN frequency collisions occur 
        \cite{hertzberg2020laserannealing,PhysRevA.101.052308}: at $\Delta_{CT} / E_C = -1,0,0.5,1$.
		In (a)-(e) data for one chip of each of the three latest processor generations is shown \cite{IBMQuantumCloud}: Montreal (``Falcon''), Manhattan (``Hummingbird'') and Washington (``Eagle''). 
        The overall frequency spread is found to be consistent with a Gaussian, although each individual $\nu$ is specifically tuned to a desired value with high accuracy.
        In (g), the $\Delta_{CT}$ distribution for untuned transmons (blue) \cite{zhang2020highfidelity} is compared to the IBM cloud chips (gray, 9 Falcons, 2 Hummingbirds) \cite{IBMQuantumCloud}. 
        There are significantly fewer collisions on the cloud devices compared to the untuned transmons, as a result of the LASIQ adjustments.
        (h) and (i) show the corresponding $\nu$ distributions which --- despite the substantial differences in $\Delta_{CT}$ --- are well described by Gaussians of similar width. 
        In (j) and (k), data for `pattern tuned' processors is shown, the `A-B Falcon' from Ref. \cite{zhang2020highfidelity} that realizes an approximate A-B pattern (ochre), and the optimal `A-B-A-C' pattern (red) \cite{hertzberg2020laserannealing}.
       These are the only configurations that show clear peaks in both the distribution for $\Delta_{CT}$ and $\nu$.
        The bin width is 50 MHz for the right column. For the left column, the bin width is 1/22 for the IBM cloud chips in panel (c),(e),(g) and 1/11 otherwise.
 		}
    \label{fig:disorder-pattern-engineered}
\end{figure}

In its current generation of cloud devices IBM employs LASIQ  to  enhance  {\em anticorrelations}, $\Delta_{CT} = \nu_C - \nu_T$,  where
$\nu_C$ and $\nu_T$ are the qubit frequencies of nearest-neighbor transmons in
the heavy-hexagon lattice geometry (denoted as ``control" and ``target" qubits). An
overview of three devices of the current 27-qubit ``Falcon", 65-qubit
``Hummingbird" and 127-qubit ``Eagle" generations is shown in
Fig.~\ref{fig:disorder-pattern-engineered}.  IBM has indeed succeeded in
imprinting some of the desired anticorrelations (left column). However,  the overall spread of qubit frequencies remains 
essentially Gaussian (right column). This being so,  we expect that the results obtained in the main text hold including for these frequency
engineered devices.

One may push  the LASIQ technique to a next level by  imprinting regular frequency {\em patterns} such as A-B or A-B-A-C into the heavy-hexagon lattice geometry currently used in all of IBM's cloud devices \cite{IBMQuantumCloud}. In this way, frequency crowding can be avoided \cite{chamberland2020, hertzberg2020laserannealing} (lowest panel of Fig.~\ref{fig:disorder-pattern-engineered}). 
Removing almost all random variations of the Josephson energy $E_J$, such layouts would implement the design philosophy  {\sc a} of the current manuscript in its purest form. As we will demonstrate below, a perfect realization of such a device would also remove the protective effects of many-body localization, and in this way compromise the integrity of quantum information.

\begin{figure}[t!]
    \centering
   \includegraphics{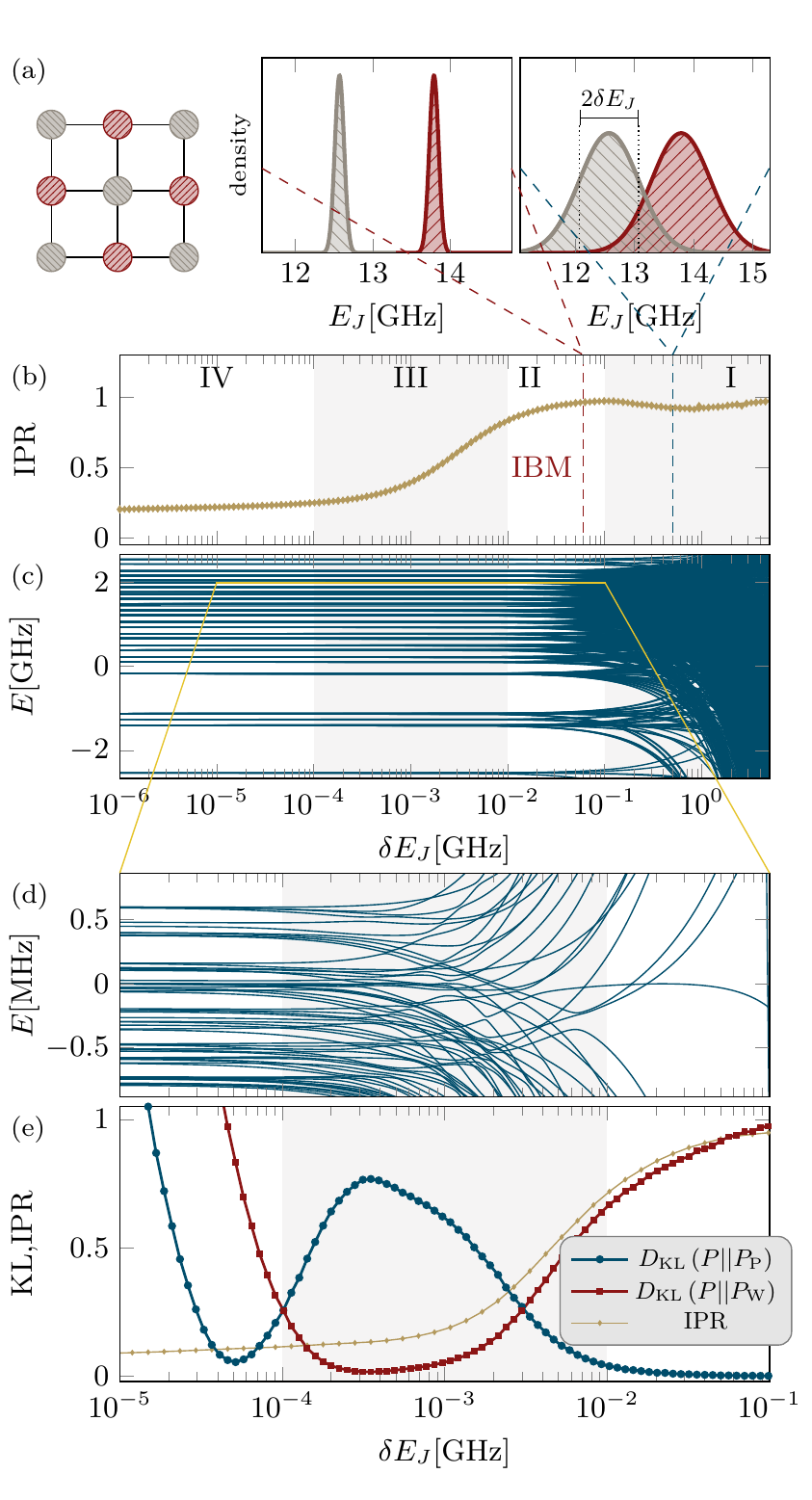}
    \caption{{\bf Effect of qubit frequency engineering.}
         (a) Staggered A-B sublattice arrangement of Josephson energies $E_J$ in a
            $3\times3$ transmon layout subject to an adjustable disorder strength
            $\delta E_J$ (right). (b) Inverse participation ratio versus $\delta E_J$. (c) Evolution of the 5-excitation bundle (with 1287 states) as a function of  $\delta E_J$ \cite{footnorm}.
            (d) Zooming into the single multiplet containing the 60 computational states defined by
            all permutations of three A transmons in  $|1\rangle$ and two in
            $|0\rangle$, two B transmons in $|1\rangle$ and two in $|0\rangle$. (e)
            Spectral statistics of  that  multiplet quantified
            by the Kullback-Leibler divergence relative to 
            Poisson and Wigner-Dyson distributions as in Fig.~\ref{fig:KL-1Dcut}, and
            the inverse participation ratio (likewise computed for states inside the
            multiplet).  The results in (b) and (e) are averaged over at least 8000 disorder
            realizations. Parameters were taken from Ref. \cite{hertzberg2020laserannealing}: on A sites (gray), we fix a mean Josephson energy $E_{J,A}=12.58~$GHz, on $B$ sites (red) $E_{J,B}=13.80~$GHz. $E_C$ is fixed to 0.33 GHz and $T=3$ MHz, a typical low value of coupling favored in current experiments.}
    \label{fig:disorder-engineering}
\end{figure}

\textbf{Pattern engineering. } 
To quantitatively discuss the role of MBL physics in the presence of  engineered qubit frequencies, we consider an A-B sublattice pattern superimposed onto the $3\times 3$ transmon lattice shown in Fig.~\ref{fig:disorder-engineering}(a). 
NN qubits are well separated in frequency, by an amount that has been deemed optimal for the operation of the cross-resonance scheme 
for performing entangling gate operations \cite{Gambetta}.
However, we note that precision engineering of alternating frequencies has a potentially problematic side effect: 
In a perfectly realized $\dots$ -A-B-A- $\dots$ arrangement, weak but finite effective next nearest neighbor coupling between degenerate A (and B)
transmons would lead to Bloch-band eigenstates. While this type of delocalization is not due to quantum chaos, chaos is in its wake the moment the inevitable presence of  residual disorder is taken into account; in what amounts to a ``fight fire with fire principle'', the  localizing counter effects of yet stronger  disorder are required to stabilize the system.

To substantiate this picture, we apply our diagnostic tools to the $3\times 3$ reference system, as summarized in Fig.~\ref{fig:disorder-engineering}. 
We identify four regimes (I-IV)  
distinguished by the value of random frequency detuning:

\begin{itemize}
	\item[I.]  Global MBL phase for $\delta E_J > 0.1~\text{GHz}$.
	\item[II.] Restructuring of the Hilbert space into energetically separated multiplets for $0.01~\text{GHz} \lesssim \delta E_J \lesssim 0.1~\text{GHz}$.
	\item[III.] Delocalization within these multiplet spaces for $10^{-4}\mathrm{GHz} \lesssim \delta E_J \lesssim 0.01$ GHz.
	\item[IV.] Restructuring of the Hilbert space into `molecular multiplets', reflecting the point group symmetries, for $\delta E_J < 10^{-4}~$GHz.
\end{itemize}

For very strong disorder, regime I, we have global MBL in a strongly coupled Fock space, as evidenced by the IPR approaching unity in Fig.~\ref{fig:disorder-engineering}(b), and the tangle of levels in Fig.~\ref{fig:disorder-engineering}(c). 

The energetically separated bundle structures that emerge in regime II, see panels (c) and (d), represent `permutation multiplets', defined by  a specific excitation structure on the two sublattices. The zoom in of  Fig.~\ref{fig:disorder-engineering}(d) shows the levels associated to a single multiplet (three A transmons in  $|1\rangle$ and two in  $|0\rangle$,  two B transmons in $|1\rangle$ and two in $|0\rangle$, any permutation). There are 60 states in this particular permutation multiplet; note that they are all computational states. In this regime II -- which is the one IBM operates in -- states are efficiently localized \emph{within} individual multiplets; the IPR is at a favorable value close to unity. 

On the blown-up $\delta E_J$ scale in (d), we see that levels are still strongly dispersing when the disorder is further reduced, 
but it is obvious by eye that we are now in a different 
situation, regime III, that is characterized by strong level repulsion. It is here that the IPR is dropping rapidly, meaning that the eigenstates are typically superpositions of states {\em within} this multiplet. 
This qualitative view is confirmed by our KL diagnostic in (e) showing that, in this region, the $R_n$ distribution strongly resembles a Wigner-Dyson distribution -- in short, we have entered a region of developed quantum chaos, a no-go area for quantum computation. {Throughout this region, our Walsh-transform diagnostic is difficult to apply, because the assignment to definite computational states is ambiguous due to the strong multiplet mixing. This is the reason we do not show any Walsh-diagnostic results in this section.}

As we proceed to even smaller disorder, there is yet a further evolution of the eigenspectrum: in regime IV the permutation multiplet resolves itself into a new set of bundles, which are now the molecular multiplets of the clean system. Here the remaining degeneracies are those of the point group symmetry of our $3\times 3$ molecule. Reflecting the correlations introduced by these symmetries, the $R_n$ distribution obeys neither Wigner-Dyson nor Poisson statistics in this regime.

Our analysis shows that the IBM work has currently landed in a good
spot between too weak and too strong disorder. At this spot, the system is clean enough to preserve the identity of small sized (computational) subspaces, yet dirty enough to achieve state-localization within these spaces. 
We expect 
the insights obtained from the quantitative discussion above to be applicable to the more elaborated A-B-A-C scheme on the heavy-hexagon lattice: Despite the lower connectivity and the more complicated pattern, NNN qubits --- the control qubits --- remain degenerate.


\bibliography{transmons}

\begin{thebibliography}{38}%
\makeatletter
\providecommand \@ifxundefined [1]{%
 \@ifx{#1\undefined}
}%
\providecommand \@ifnum [1]{%
 \ifnum #1\expandafter \@firstoftwo
 \else \expandafter \@secondoftwo
 \fi
}%
\providecommand \@ifx [1]{%
 \ifx #1\expandafter \@firstoftwo
 \else \expandafter \@secondoftwo
 \fi
}%
\providecommand \natexlab [1]{#1}%
\providecommand \enquote  [1]{``#1''}%
\providecommand \bibnamefont  [1]{#1}%
\providecommand \bibfnamefont [1]{#1}%
\providecommand \citenamefont [1]{#1}%
\providecommand \href@noop [0]{\@secondoftwo}%
\providecommand \href [0]{\begingroup \@sanitize@url \@href}%
\providecommand \@href[1]{\@@startlink{#1}\@@href}%
\providecommand \@@href[1]{\endgroup#1\@@endlink}%
\providecommand \@sanitize@url [0]{\catcode `\\12\catcode `\$12\catcode
  `\&12\catcode `\#12\catcode `\^12\catcode `\_12\catcode `\%12\relax}%
\providecommand \@@startlink[1]{}%
\providecommand \@@endlink[0]{}%
\providecommand \url  [0]{\begingroup\@sanitize@url \@url }%
\providecommand \@url [1]{\endgroup\@href {#1}{\urlprefix }}%
\providecommand \urlprefix  [0]{URL }%
\providecommand \Eprint [0]{\href }%
\providecommand \doibase [0]{https://doi.org/}%
\providecommand \selectlanguage [0]{\@gobble}%
\providecommand \bibinfo  [0]{\@secondoftwo}%
\providecommand \bibfield  [0]{\@secondoftwo}%
\providecommand \translation [1]{[#1]}%
\providecommand \BibitemOpen [0]{}%
\providecommand \bibitemStop [0]{}%
\providecommand \bibitemNoStop [0]{.\EOS\space}%
\providecommand \EOS [0]{\spacefactor3000\relax}%
\providecommand \BibitemShut  [1]{\csname bibitem#1\endcsname}%
\let\auto@bib@innerbib\@empty
\bibitem [{Note1()}]{Note1}%
  \BibitemOpen
  \bibinfo {note} {In the MBL literature these are called $p$-bits and
  $l$-bits; we modify the notation to connect to modern quantum-information
  usage.}\BibitemShut {Stop}%
\bibitem [{\citenamefont {Huse}\ \emph {et~al.}(2014)\citenamefont {Huse},
  \citenamefont {Nandkishore},\ and\ \citenamefont
  {Oganesyan}}]{PhysRevB.90.174202}%
  \BibitemOpen
  \bibfield  {author} {\bibinfo {author} {\bibfnamefont {D.~A.}\ \bibnamefont
  {Huse}}, \bibinfo {author} {\bibfnamefont {R.}~\bibnamefont {Nandkishore}},\
  and\ \bibinfo {author} {\bibfnamefont {V.}~\bibnamefont {Oganesyan}},\
  }\bibfield  {title} {\bibinfo {title} {{Phenomenology of fully
  many-body-localized systems}},\ }\href
  {https://doi.org/10.1103/PhysRevB.90.174202} {\bibfield  {journal} {\bibinfo
  {journal} {Phys. Rev. B}\ }\textbf {\bibinfo {volume} {90}},\ \bibinfo
  {pages} {174202} (\bibinfo {year} {2014})}\BibitemShut {NoStop}%
\bibitem [{\citenamefont {Serbyn}\ \emph {et~al.}(2013)\citenamefont {Serbyn},
  \citenamefont {Papi\ifmmode~\acute{c}\else \'{c}\fi{}},\ and\ \citenamefont
  {Abanin}}]{PhysRevLett.111.127201}%
  \BibitemOpen
  \bibfield  {author} {\bibinfo {author} {\bibfnamefont {M.}~\bibnamefont
  {Serbyn}}, \bibinfo {author} {\bibfnamefont {Z.}~\bibnamefont
  {Papi\ifmmode~\acute{c}\else \'{c}\fi{}}},\ and\ \bibinfo {author}
  {\bibfnamefont {D.~A.}\ \bibnamefont {Abanin}},\ }\bibfield  {title}
  {\bibinfo {title} {{Local Conservation Laws and the Structure of the
  Many-Body Localized States}},\ }\href
  {https://doi.org/10.1103/PhysRevLett.111.127201} {\bibfield  {journal}
  {\bibinfo  {journal} {Phys. Rev. Lett.}\ }\textbf {\bibinfo {volume} {111}},\
  \bibinfo {pages} {127201} (\bibinfo {year} {2013})}\BibitemShut {NoStop}%
\bibitem [{\citenamefont {Ku}\ \emph {et~al.}(2020)\citenamefont {Ku},
  \citenamefont {Xu}, \citenamefont {Brink}, \citenamefont {McKay},
  \citenamefont {Hertzberg}, \citenamefont {Ansari},\ and\ \citenamefont
  {Plourde}}]{PhysRevLett.125.200504}%
  \BibitemOpen
  \bibfield  {author} {\bibinfo {author} {\bibfnamefont {J.}~\bibnamefont
  {Ku}}, \bibinfo {author} {\bibfnamefont {X.}~\bibnamefont {Xu}}, \bibinfo
  {author} {\bibfnamefont {M.}~\bibnamefont {Brink}}, \bibinfo {author}
  {\bibfnamefont {D.~C.}\ \bibnamefont {McKay}}, \bibinfo {author}
  {\bibfnamefont {J.~B.}\ \bibnamefont {Hertzberg}}, \bibinfo {author}
  {\bibfnamefont {M.~H.}\ \bibnamefont {Ansari}},\ and\ \bibinfo {author}
  {\bibfnamefont {B.~L.~T.}\ \bibnamefont {Plourde}},\ }\bibfield  {title}
  {\bibinfo {title} {{Suppression of Unwanted {\em ZZ} Interactions in a Hybrid
  Two-Qubit System}},\ }\href {https://doi.org/10.1103/PhysRevLett.125.200504}
  {\bibfield  {journal} {\bibinfo  {journal} {Phys. Rev. Lett.}\ }\textbf
  {\bibinfo {volume} {125}},\ \bibinfo {pages} {200504} (\bibinfo {year}
  {2020})}\BibitemShut {NoStop}%
\bibitem [{\citenamefont {Orell}\ \emph {et~al.}(2019)\citenamefont {Orell},
  \citenamefont {Michailidis}, \citenamefont {Serbyn},\ and\ \citenamefont
  {Silveri}}]{orell_probing_2019}%
  \BibitemOpen
  \bibfield  {author} {\bibinfo {author} {\bibfnamefont {T.}~\bibnamefont
  {Orell}}, \bibinfo {author} {\bibfnamefont {A.~A.}\ \bibnamefont
  {Michailidis}}, \bibinfo {author} {\bibfnamefont {M.}~\bibnamefont
  {Serbyn}},\ and\ \bibinfo {author} {\bibfnamefont {M.}~\bibnamefont
  {Silveri}},\ }\bibfield  {title} {\bibinfo {title} {{Probing the many-body
  localization phase transition with superconducting circuits}},\ }\href
  {https://doi.org/10.1103/PhysRevB.100.134504} {\bibfield  {journal} {\bibinfo
   {journal} {Phys. Rev. B}\ }\textbf {\bibinfo {volume} {100}},\ \bibinfo
  {pages} {134504} (\bibinfo {year} {2019})}\BibitemShut {NoStop}%
\bibitem [{\citenamefont {Throckmorton}\ and\ \citenamefont
  {Das~Sarma}(2021)}]{Throckmorton2020}%
  \BibitemOpen
  \bibfield  {author} {\bibinfo {author} {\bibfnamefont {R.~E.}\ \bibnamefont
  {Throckmorton}}\ and\ \bibinfo {author} {\bibfnamefont {S.}~\bibnamefont
  {Das~Sarma}},\ }\bibfield  {title} {\bibinfo {title} {Studying many-body
  localization in exchange-coupled electron spin qubits using spin-spin
  correlations},\ }\href {https://doi.org/10.1103/PhysRevB.103.165431}
  {\bibfield  {journal} {\bibinfo  {journal} {Phys. Rev. B}\ }\textbf {\bibinfo
  {volume} {103}},\ \bibinfo {pages} {165431} (\bibinfo {year}
  {2021})}\BibitemShut {NoStop}%
\bibitem [{Note2()}]{Note2}%
  \BibitemOpen
  \bibinfo {note} {We may incidentally remark that this work started from a
  project study~\cite {borner_simon-dominik_classical_2020} on \protect \emph
  {classical} chaos in the transmon system. Classically, the transmon
  Hamiltonian Eq.~\protect \textup {\hbox {\mathsurround \z@ \protect
  \normalfont (\ignorespaces \ref {trans1}\unskip \@@italiccorr )}} describes a
  system of coupled mathematical pendula of mass $m=1/8E_C$ and gravitational
  acceleration $g=8E_C E_J$ (in units where $\hbar =1$ and $\ell =1$ (pendulum
  length)). Nonlinearly coupled pendula generally show a transition from
  integrable motion at low energies to hard chaos at high energies. (There are
  desktop gimmicks with just two coupled masses demonstrating the phenomenon.)
  The principal observation of the project was that already the classical two
  transmon Hamiltonian showed tendencies to chaos when excited to sufficiently
  high energies. The generalization to ten coupled oscillators made the
  situation worse, with Lyapunov exponents signaling uncontrollable dynamics
  for energies matching those of QC applications with $0$ and $1$ qubit states,
  and at time scales way below typical coherence times.}\BibitemShut {Stop}%
\bibitem [{\citenamefont {Versluis}\ \emph {et~al.}(2017)\citenamefont
  {Versluis}, \citenamefont {Poletto}, \citenamefont {Khammassi}, \citenamefont
  {Tarasinski}, \citenamefont {Haider}, \citenamefont {Michalak}, \citenamefont
  {Bruno}, \citenamefont {Bertels},\ and\ \citenamefont
  {DiCarlo}}]{PhysRevApplied.8.034021}%
  \BibitemOpen
  \bibfield  {author} {\bibinfo {author} {\bibfnamefont {R.}~\bibnamefont
  {Versluis}}, \bibinfo {author} {\bibfnamefont {S.}~\bibnamefont {Poletto}},
  \bibinfo {author} {\bibfnamefont {N.}~\bibnamefont {Khammassi}}, \bibinfo
  {author} {\bibfnamefont {B.}~\bibnamefont {Tarasinski}}, \bibinfo {author}
  {\bibfnamefont {N.}~\bibnamefont {Haider}}, \bibinfo {author} {\bibfnamefont
  {D.~J.}\ \bibnamefont {Michalak}}, \bibinfo {author} {\bibfnamefont
  {A.}~\bibnamefont {Bruno}}, \bibinfo {author} {\bibfnamefont
  {K.}~\bibnamefont {Bertels}},\ and\ \bibinfo {author} {\bibfnamefont
  {L.}~\bibnamefont {DiCarlo}},\ }\bibfield  {title} {\bibinfo {title}
  {{Scalable Quantum Circuit and Control for a Superconducting Surface Code}},\
  }\href {https://doi.org/10.1103/PhysRevApplied.8.034021} {\bibfield
  {journal} {\bibinfo  {journal} {Phys. Rev. Applied}\ }\textbf {\bibinfo
  {volume} {8}},\ \bibinfo {pages} {034021} (\bibinfo {year}
  {2017})}\BibitemShut {NoStop}%
\bibitem [{\citenamefont {Arute}\ \emph {et~al.}(2019)\citenamefont {Arute},
  \citenamefont {Arya}, \citenamefont {Babbush}, \citenamefont {Bacon},
  \citenamefont {Bardin}, \citenamefont {Barends}, \citenamefont {Biswas},
  \citenamefont {Boixo}, \citenamefont {Brandao}, \citenamefont {Buell},
  \citenamefont {Burkett}, \citenamefont {Chen}, \citenamefont {Chen},
  \citenamefont {Chiaro}, \citenamefont {Collins}, \citenamefont {Courtney},
  \citenamefont {Dunsworth}, \citenamefont {Farhi}, \citenamefont {Foxen},
  \citenamefont {Fowler}, \citenamefont {Gidney}, \citenamefont {Giustina},
  \citenamefont {Graff}, \citenamefont {Guerin}, \citenamefont {Habegger},
  \citenamefont {Harrigan}, \citenamefont {Hartmann}, \citenamefont {Ho},
  \citenamefont {Hoffmann}, \citenamefont {Huang}, \citenamefont {Humble},
  \citenamefont {Isakov}, \citenamefont {Jeffrey}, \citenamefont {Jiang},
  \citenamefont {Kafri}, \citenamefont {Kechedzhi}, \citenamefont {Kelly},
  \citenamefont {Klimov}, \citenamefont {Knysh}, \citenamefont {Korotkov},
  \citenamefont {Kostritsa}, \citenamefont {Landhuis}, \citenamefont
  {Lindmark}, \citenamefont {Lucero}, \citenamefont {Lyakh}, \citenamefont
  {Mandr{\`a}}, \citenamefont {McClean}, \citenamefont {McEwen}, \citenamefont
  {Megrant}, \citenamefont {Mi}, \citenamefont {Michielsen}, \citenamefont
  {Mohseni}, \citenamefont {Mutus}, \citenamefont {Naaman}, \citenamefont
  {Neeley}, \citenamefont {Neill}, \citenamefont {Niu}, \citenamefont {Ostby},
  \citenamefont {Petukhov}, \citenamefont {Platt}, \citenamefont {Quintana},
  \citenamefont {Rieffel}, \citenamefont {Roushan}, \citenamefont {Rubin},
  \citenamefont {Sank}, \citenamefont {Satzinger}, \citenamefont {Smelyanskiy},
  \citenamefont {Sung}, \citenamefont {Trevithick}, \citenamefont
  {Vainsencher}, \citenamefont {Villalonga}, \citenamefont {White},
  \citenamefont {Yao}, \citenamefont {Yeh}, \citenamefont {Zalcman},
  \citenamefont {Neven},\ and\ \citenamefont
  {Martinis}}]{GoogleQuantumSupremacy}%
  \BibitemOpen
  \bibfield  {author} {\bibinfo {author} {\bibfnamefont {F.}~\bibnamefont
  {Arute}}, \bibinfo {author} {\bibfnamefont {K.}~\bibnamefont {Arya}},
  \bibinfo {author} {\bibfnamefont {R.}~\bibnamefont {Babbush}}, \bibinfo
  {author} {\bibfnamefont {D.}~\bibnamefont {Bacon}}, \bibinfo {author}
  {\bibfnamefont {J.~C.}\ \bibnamefont {Bardin}}, \bibinfo {author}
  {\bibfnamefont {R.}~\bibnamefont {Barends}}, \bibinfo {author} {\bibfnamefont
  {R.}~\bibnamefont {Biswas}}, \bibinfo {author} {\bibfnamefont
  {S.}~\bibnamefont {Boixo}}, \bibinfo {author} {\bibfnamefont {F.~G. S.~L.}\
  \bibnamefont {Brandao}}, \bibinfo {author} {\bibfnamefont {D.~A.}\
  \bibnamefont {Buell}}, \bibinfo {author} {\bibfnamefont {B.}~\bibnamefont
  {Burkett}}, \bibinfo {author} {\bibfnamefont {Y.}~\bibnamefont {Chen}},
  \bibinfo {author} {\bibfnamefont {Z.}~\bibnamefont {Chen}}, \bibinfo {author}
  {\bibfnamefont {B.}~\bibnamefont {Chiaro}}, \bibinfo {author} {\bibfnamefont
  {R.}~\bibnamefont {Collins}}, \bibinfo {author} {\bibfnamefont
  {W.}~\bibnamefont {Courtney}}, \bibinfo {author} {\bibfnamefont
  {A.}~\bibnamefont {Dunsworth}}, \bibinfo {author} {\bibfnamefont
  {E.}~\bibnamefont {Farhi}}, \bibinfo {author} {\bibfnamefont
  {B.}~\bibnamefont {Foxen}}, \bibinfo {author} {\bibfnamefont
  {A.}~\bibnamefont {Fowler}}, \bibinfo {author} {\bibfnamefont
  {C.}~\bibnamefont {Gidney}}, \bibinfo {author} {\bibfnamefont
  {M.}~\bibnamefont {Giustina}}, \bibinfo {author} {\bibfnamefont
  {R.}~\bibnamefont {Graff}}, \bibinfo {author} {\bibfnamefont
  {K.}~\bibnamefont {Guerin}}, \bibinfo {author} {\bibfnamefont
  {S.}~\bibnamefont {Habegger}}, \bibinfo {author} {\bibfnamefont {M.~P.}\
  \bibnamefont {Harrigan}}, \bibinfo {author} {\bibfnamefont {M.~J.}\
  \bibnamefont {Hartmann}}, \bibinfo {author} {\bibfnamefont {A.}~\bibnamefont
  {Ho}}, \bibinfo {author} {\bibfnamefont {M.}~\bibnamefont {Hoffmann}},
  \bibinfo {author} {\bibfnamefont {T.}~\bibnamefont {Huang}}, \bibinfo
  {author} {\bibfnamefont {T.~S.}\ \bibnamefont {Humble}}, \bibinfo {author}
  {\bibfnamefont {S.~V.}\ \bibnamefont {Isakov}}, \bibinfo {author}
  {\bibfnamefont {E.}~\bibnamefont {Jeffrey}}, \bibinfo {author} {\bibfnamefont
  {Z.}~\bibnamefont {Jiang}}, \bibinfo {author} {\bibfnamefont
  {D.}~\bibnamefont {Kafri}}, \bibinfo {author} {\bibfnamefont
  {K.}~\bibnamefont {Kechedzhi}}, \bibinfo {author} {\bibfnamefont
  {J.}~\bibnamefont {Kelly}}, \bibinfo {author} {\bibfnamefont {P.~V.}\
  \bibnamefont {Klimov}}, \bibinfo {author} {\bibfnamefont {S.}~\bibnamefont
  {Knysh}}, \bibinfo {author} {\bibfnamefont {A.}~\bibnamefont {Korotkov}},
  \bibinfo {author} {\bibfnamefont {F.}~\bibnamefont {Kostritsa}}, \bibinfo
  {author} {\bibfnamefont {D.}~\bibnamefont {Landhuis}}, \bibinfo {author}
  {\bibfnamefont {M.}~\bibnamefont {Lindmark}}, \bibinfo {author}
  {\bibfnamefont {E.}~\bibnamefont {Lucero}}, \bibinfo {author} {\bibfnamefont
  {D.}~\bibnamefont {Lyakh}}, \bibinfo {author} {\bibfnamefont
  {S.}~\bibnamefont {Mandr{\`a}}}, \bibinfo {author} {\bibfnamefont {J.~R.}\
  \bibnamefont {McClean}}, \bibinfo {author} {\bibfnamefont {M.}~\bibnamefont
  {McEwen}}, \bibinfo {author} {\bibfnamefont {A.}~\bibnamefont {Megrant}},
  \bibinfo {author} {\bibfnamefont {X.}~\bibnamefont {Mi}}, \bibinfo {author}
  {\bibfnamefont {K.}~\bibnamefont {Michielsen}}, \bibinfo {author}
  {\bibfnamefont {M.}~\bibnamefont {Mohseni}}, \bibinfo {author} {\bibfnamefont
  {J.}~\bibnamefont {Mutus}}, \bibinfo {author} {\bibfnamefont
  {O.}~\bibnamefont {Naaman}}, \bibinfo {author} {\bibfnamefont
  {M.}~\bibnamefont {Neeley}}, \bibinfo {author} {\bibfnamefont
  {C.}~\bibnamefont {Neill}}, \bibinfo {author} {\bibfnamefont {M.~Y.}\
  \bibnamefont {Niu}}, \bibinfo {author} {\bibfnamefont {E.}~\bibnamefont
  {Ostby}}, \bibinfo {author} {\bibfnamefont {A.}~\bibnamefont {Petukhov}},
  \bibinfo {author} {\bibfnamefont {J.~C.}\ \bibnamefont {Platt}}, \bibinfo
  {author} {\bibfnamefont {C.}~\bibnamefont {Quintana}}, \bibinfo {author}
  {\bibfnamefont {E.~G.}\ \bibnamefont {Rieffel}}, \bibinfo {author}
  {\bibfnamefont {P.}~\bibnamefont {Roushan}}, \bibinfo {author} {\bibfnamefont
  {N.~C.}\ \bibnamefont {Rubin}}, \bibinfo {author} {\bibfnamefont
  {D.}~\bibnamefont {Sank}}, \bibinfo {author} {\bibfnamefont {K.~J.}\
  \bibnamefont {Satzinger}}, \bibinfo {author} {\bibfnamefont {V.}~\bibnamefont
  {Smelyanskiy}}, \bibinfo {author} {\bibfnamefont {K.~J.}\ \bibnamefont
  {Sung}}, \bibinfo {author} {\bibfnamefont {M.~D.}\ \bibnamefont
  {Trevithick}}, \bibinfo {author} {\bibfnamefont {A.}~\bibnamefont
  {Vainsencher}}, \bibinfo {author} {\bibfnamefont {B.}~\bibnamefont
  {Villalonga}}, \bibinfo {author} {\bibfnamefont {T.}~\bibnamefont {White}},
  \bibinfo {author} {\bibfnamefont {Z.~J.}\ \bibnamefont {Yao}}, \bibinfo
  {author} {\bibfnamefont {P.}~\bibnamefont {Yeh}}, \bibinfo {author}
  {\bibfnamefont {A.}~\bibnamefont {Zalcman}}, \bibinfo {author} {\bibfnamefont
  {H.}~\bibnamefont {Neven}},\ and\ \bibinfo {author} {\bibfnamefont {J.~M.}\
  \bibnamefont {Martinis}},\ }\bibfield  {title} {\bibinfo {title} {{Quantum
  supremacy using a programmable superconducting processor}},\ }\href
  {https://doi.org/10.1038/s41586-019-1666-5} {\bibfield  {journal} {\bibinfo
  {journal} {Nature}\ }\textbf {\bibinfo {volume} {574}},\ \bibinfo {pages}
  {505} (\bibinfo {year} {2019})}\BibitemShut {NoStop}%
\bibitem [{\citenamefont {{Córcoles}}\ \emph {et~al.}(2020)\citenamefont
  {{Córcoles}}, \citenamefont {{Kandala}}, \citenamefont {{Javadi-Abhari}},
  \citenamefont {{McClure}}, \citenamefont {{Cross}}, \citenamefont {{Temme}},
  \citenamefont {{Nation}}, \citenamefont {{Steffen}},\ and\ \citenamefont
  {{Gambetta}}}]{8936946}%
  \BibitemOpen
  \bibfield  {author} {\bibinfo {author} {\bibfnamefont {A.~D.}\ \bibnamefont
  {{Córcoles}}}, \bibinfo {author} {\bibfnamefont {A.}~\bibnamefont
  {{Kandala}}}, \bibinfo {author} {\bibfnamefont {A.}~\bibnamefont
  {{Javadi-Abhari}}}, \bibinfo {author} {\bibfnamefont {D.~T.}\ \bibnamefont
  {{McClure}}}, \bibinfo {author} {\bibfnamefont {A.~W.}\ \bibnamefont
  {{Cross}}}, \bibinfo {author} {\bibfnamefont {K.}~\bibnamefont {{Temme}}},
  \bibinfo {author} {\bibfnamefont {P.~D.}\ \bibnamefont {{Nation}}}, \bibinfo
  {author} {\bibfnamefont {M.}~\bibnamefont {{Steffen}}},\ and\ \bibinfo
  {author} {\bibfnamefont {J.~M.}\ \bibnamefont {{Gambetta}}},\ }\bibfield
  {title} {\bibinfo {title} {{Challenges and Opportunities of Near-Term Quantum
  Computing Systems}},\ }\href {https://doi.org/10.1109/JPROC.2019.2954005}
  {\bibfield  {journal} {\bibinfo  {journal} {Proceedings of the IEEE}\
  }\textbf {\bibinfo {volume} {108}},\ \bibinfo {pages} {1338} (\bibinfo {year}
  {2020})}\BibitemShut {NoStop}%
\bibitem [{\citenamefont {Potter}\ \emph {et~al.}(2015)\citenamefont {Potter},
  \citenamefont {Vasseur},\ and\ \citenamefont
  {Parameswaran}}]{PhysRevX.5.031033}%
  \BibitemOpen
  \bibfield  {author} {\bibinfo {author} {\bibfnamefont {A.~C.}\ \bibnamefont
  {Potter}}, \bibinfo {author} {\bibfnamefont {R.}~\bibnamefont {Vasseur}},\
  and\ \bibinfo {author} {\bibfnamefont {S.~A.}\ \bibnamefont {Parameswaran}},\
  }\bibfield  {title} {\bibinfo {title} {{Universal Properties of Many-Body
  Delocalization Transitions}},\ }\href
  {https://doi.org/10.1103/PhysRevX.5.031033} {\bibfield  {journal} {\bibinfo
  {journal} {Phys. Rev. X}\ }\textbf {\bibinfo {volume} {5}},\ \bibinfo {pages}
  {031033} (\bibinfo {year} {2015})}\BibitemShut {NoStop}%
\bibitem [{\citenamefont {Koch}\ \emph {et~al.}(2007)\citenamefont {Koch},
  \citenamefont {Yu}, \citenamefont {Gambetta}, \citenamefont {Houck},
  \citenamefont {Schuster}, \citenamefont {Majer}, \citenamefont {Blais},
  \citenamefont {Devoret}, \citenamefont {Girvin},\ and\ \citenamefont
  {Schoelkopf}}]{koch_charge-insensitive_2007}%
  \BibitemOpen
  \bibfield  {author} {\bibinfo {author} {\bibfnamefont {J.}~\bibnamefont
  {Koch}}, \bibinfo {author} {\bibfnamefont {T.~M.}\ \bibnamefont {Yu}},
  \bibinfo {author} {\bibfnamefont {J.}~\bibnamefont {Gambetta}}, \bibinfo
  {author} {\bibfnamefont {A.~A.}\ \bibnamefont {Houck}}, \bibinfo {author}
  {\bibfnamefont {D.~I.}\ \bibnamefont {Schuster}}, \bibinfo {author}
  {\bibfnamefont {J.}~\bibnamefont {Majer}}, \bibinfo {author} {\bibfnamefont
  {A.}~\bibnamefont {Blais}}, \bibinfo {author} {\bibfnamefont {M.~H.}\
  \bibnamefont {Devoret}}, \bibinfo {author} {\bibfnamefont {S.~M.}\
  \bibnamefont {Girvin}},\ and\ \bibinfo {author} {\bibfnamefont {R.~J.}\
  \bibnamefont {Schoelkopf}},\ }\bibfield  {title} {\bibinfo {title}
  {{Charge-insensitive qubit design derived from the Cooper pair box}},\ }\href
  {https://doi.org/10.1103/PhysRevA.76.042319} {\bibfield  {journal} {\bibinfo
  {journal} {Phys. Rev. A}\ }\textbf {\bibinfo {volume} {76}},\ \bibinfo
  {pages} {042319} (\bibinfo {year} {2007})}\BibitemShut {NoStop}%
\bibitem [{\citenamefont {{Gambetta}}(2013)}]{Gambetta}%
  \BibitemOpen
  \bibfield  {author} {\bibinfo {author} {\bibfnamefont {J.~M.}\ \bibnamefont
  {{Gambetta}}},\ }\bibfield  {title} {\bibinfo {title} {{Control of
  Superconducting Qubits}},\ }\href
  {https://www.fzj.de/SharedDocs/Downloads/PGI/EN/SpringSchool/Lecture-Notes-Book-Form/Skriptbuch-2013.pdf}
  {\bibfield  {journal} {\bibinfo  {journal} {in, Proceedings of the 44th IFF
  Spring School, ``Quantum Information Processing'', Forschungszentrum
  J{\"u}lich}\ } (\bibinfo {year} {2013})}\BibitemShut {NoStop}%
\bibitem [{\citenamefont {Blais}\ \emph {et~al.}(2021)\citenamefont {Blais},
  \citenamefont {Grimsmo}, \citenamefont {Girvin},\ and\ \citenamefont
  {Wallraff}}]{blais_circuit_2020}%
  \BibitemOpen
  \bibfield  {author} {\bibinfo {author} {\bibfnamefont {A.}~\bibnamefont
  {Blais}}, \bibinfo {author} {\bibfnamefont {A.~L.}\ \bibnamefont {Grimsmo}},
  \bibinfo {author} {\bibfnamefont {S.~M.}\ \bibnamefont {Girvin}},\ and\
  \bibinfo {author} {\bibfnamefont {A.}~\bibnamefont {Wallraff}},\ }\bibfield
  {title} {\bibinfo {title} {Circuit quantum electrodynamics},\ }\href
  {https://doi.org/10.1103/RevModPhys.93.025005} {\bibfield  {journal}
  {\bibinfo  {journal} {Rev. Mod. Phys.}\ }\textbf {\bibinfo {volume} {93}},\
  \bibinfo {pages} {025005} (\bibinfo {year} {2021})}\BibitemShut {NoStop}%
\bibitem [{IBM()}]{IBMQuantumCloud}%
  \BibitemOpen
  \href {https://www.ibm.com/quantum-computing/systems/} {\bibinfo {title}
  {https://www.ibm.com/quantum-computing/}}\BibitemShut {NoStop}%
\bibitem [{\citenamefont {Hertzberg}\ \emph {et~al.}(2021)\citenamefont
  {Hertzberg}, \citenamefont {Zhang}, \citenamefont {Rosenblatt}, \citenamefont
  {Magesan}, \citenamefont {Smolin}, \citenamefont {Yau}, \citenamefont
  {Adiga}, \citenamefont {Sandberg}, \citenamefont {Brink}, \citenamefont
  {Chow},\ and\ \citenamefont {Orcutt}}]{hertzberg2020laserannealing}%
  \BibitemOpen
  \bibfield  {author} {\bibinfo {author} {\bibfnamefont {J.~B.}\ \bibnamefont
  {Hertzberg}}, \bibinfo {author} {\bibfnamefont {E.~J.}\ \bibnamefont
  {Zhang}}, \bibinfo {author} {\bibfnamefont {S.}~\bibnamefont {Rosenblatt}},
  \bibinfo {author} {\bibfnamefont {E.}~\bibnamefont {Magesan}}, \bibinfo
  {author} {\bibfnamefont {J.~A.}\ \bibnamefont {Smolin}}, \bibinfo {author}
  {\bibfnamefont {J.-B.}\ \bibnamefont {Yau}}, \bibinfo {author} {\bibfnamefont
  {V.~P.}\ \bibnamefont {Adiga}}, \bibinfo {author} {\bibfnamefont
  {M.}~\bibnamefont {Sandberg}}, \bibinfo {author} {\bibfnamefont
  {M.}~\bibnamefont {Brink}}, \bibinfo {author} {\bibfnamefont {J.~M.}\
  \bibnamefont {Chow}},\ and\ \bibinfo {author} {\bibfnamefont {J.~S.}\
  \bibnamefont {Orcutt}},\ }\bibfield  {title} {\bibinfo {title}
  {{Laser-annealing Josephson junctions for yielding scaled-up superconducting
  quantum processors}},\ }\href {https://doi.org/10.1038/s41534-021-00464-5}
  {\bibfield  {journal} {\bibinfo  {journal} {npj Quantum Information}\
  }\textbf {\bibinfo {volume} {7}},\ \bibinfo {pages} {129} (\bibinfo {year}
  {2021})}\BibitemShut {NoStop}%
\bibitem [{\citenamefont {Zhang}\ \emph {et~al.}(2020)\citenamefont {Zhang},
  \citenamefont {Srinivasan}, \citenamefont {Sundaresan}, \citenamefont
  {Bogorin}, \citenamefont {Martin}, \citenamefont {Hertzberg}, \citenamefont
  {Timmerwilke}, \citenamefont {Pritchett}, \citenamefont {Yau}, \citenamefont
  {Wang}, \citenamefont {Landers}, \citenamefont {Lewandowski}, \citenamefont
  {Narasgond}, \citenamefont {Rosenblatt}, \citenamefont {Keefe}, \citenamefont
  {Lauer}, \citenamefont {Rothwell}, \citenamefont {McClure}, \citenamefont
  {Dial}, \citenamefont {Orcutt}, \citenamefont {Brink},\ and\ \citenamefont
  {Chow}}]{zhang2020highfidelity}%
  \BibitemOpen
  \bibfield  {author} {\bibinfo {author} {\bibfnamefont {E.~J.}\ \bibnamefont
  {Zhang}}, \bibinfo {author} {\bibfnamefont {S.}~\bibnamefont {Srinivasan}},
  \bibinfo {author} {\bibfnamefont {N.}~\bibnamefont {Sundaresan}}, \bibinfo
  {author} {\bibfnamefont {D.~F.}\ \bibnamefont {Bogorin}}, \bibinfo {author}
  {\bibfnamefont {Y.}~\bibnamefont {Martin}}, \bibinfo {author} {\bibfnamefont
  {J.~B.}\ \bibnamefont {Hertzberg}}, \bibinfo {author} {\bibfnamefont
  {J.}~\bibnamefont {Timmerwilke}}, \bibinfo {author} {\bibfnamefont {E.~J.}\
  \bibnamefont {Pritchett}}, \bibinfo {author} {\bibfnamefont {J.-B.}\
  \bibnamefont {Yau}}, \bibinfo {author} {\bibfnamefont {C.}~\bibnamefont
  {Wang}}, \bibinfo {author} {\bibfnamefont {W.}~\bibnamefont {Landers}},
  \bibinfo {author} {\bibfnamefont {E.~P.}\ \bibnamefont {Lewandowski}},
  \bibinfo {author} {\bibfnamefont {A.}~\bibnamefont {Narasgond}}, \bibinfo
  {author} {\bibfnamefont {S.}~\bibnamefont {Rosenblatt}}, \bibinfo {author}
  {\bibfnamefont {G.~A.}\ \bibnamefont {Keefe}}, \bibinfo {author}
  {\bibfnamefont {I.}~\bibnamefont {Lauer}}, \bibinfo {author} {\bibfnamefont
  {M.~B.}\ \bibnamefont {Rothwell}}, \bibinfo {author} {\bibfnamefont {D.~T.}\
  \bibnamefont {McClure}}, \bibinfo {author} {\bibfnamefont {O.~E.}\
  \bibnamefont {Dial}}, \bibinfo {author} {\bibfnamefont {J.~S.}\ \bibnamefont
  {Orcutt}}, \bibinfo {author} {\bibfnamefont {M.}~\bibnamefont {Brink}},\ and\
  \bibinfo {author} {\bibfnamefont {J.~M.}\ \bibnamefont {Chow}},\ }\href@noop
  {} {\bibinfo {title} {High-fidelity superconducting quantum processors via
  laser-annealing of transmon qubits}} (\bibinfo {year} {2020}),\ \Eprint
  {https://arxiv.org/abs/2012.08475} {arXiv:2012.08475 [quant-ph]} \BibitemShut
  {NoStop}%
\bibitem [{\citenamefont {Mac\'e}\ \emph {et~al.}(2019)\citenamefont {Mac\'e},
  \citenamefont {Alet},\ and\ \citenamefont
  {Laflorencie}}]{PhysRevLett.123.180601}%
  \BibitemOpen
  \bibfield  {author} {\bibinfo {author} {\bibfnamefont {N.}~\bibnamefont
  {Mac\'e}}, \bibinfo {author} {\bibfnamefont {F.}~\bibnamefont {Alet}},\ and\
  \bibinfo {author} {\bibfnamefont {N.}~\bibnamefont {Laflorencie}},\
  }\bibfield  {title} {\bibinfo {title} {{Multifractal Scalings Across the
  Many-Body Localization Transition}},\ }\href
  {https://doi.org/10.1103/PhysRevLett.123.180601} {\bibfield  {journal}
  {\bibinfo  {journal} {Phys. Rev. Lett.}\ }\textbf {\bibinfo {volume} {123}},\
  \bibinfo {pages} {180601} (\bibinfo {year} {2019})}\BibitemShut {NoStop}%
\bibitem [{\citenamefont {Serbyn}\ and\ \citenamefont
  {Moore}(2016)}]{Serbyn2016}%
  \BibitemOpen
  \bibfield  {author} {\bibinfo {author} {\bibfnamefont {M.}~\bibnamefont
  {Serbyn}}\ and\ \bibinfo {author} {\bibfnamefont {J.~E.}\ \bibnamefont
  {Moore}},\ }\bibfield  {title} {\bibinfo {title} {Spectral statistics across
  the many-body localization transition},\ }\href
  {https://doi.org/10.1103/PhysRevB.93.041424} {\bibfield  {journal} {\bibinfo
  {journal} {Phys. Rev. B}\ }\textbf {\bibinfo {volume} {93}},\ \bibinfo
  {pages} {041424} (\bibinfo {year} {2016})}\BibitemShut {NoStop}%
\bibitem [{Note3()}]{Note3}%
  \BibitemOpen
  \bibinfo {note} {For the $N=10$ transmon chain at hand, this manifold
  contains a total of 2002 different states.}\BibitemShut {Stop}%
\bibitem [{\citenamefont {Evers}\ and\ \citenamefont
  {Mirlin}(2008)}]{RevModPhys.80.1355}%
  \BibitemOpen
  \bibfield  {author} {\bibinfo {author} {\bibfnamefont {F.}~\bibnamefont
  {Evers}}\ and\ \bibinfo {author} {\bibfnamefont {A.~D.}\ \bibnamefont
  {Mirlin}},\ }\bibfield  {title} {\bibinfo {title} {Anderson transitions},\
  }\href {https://doi.org/10.1103/RevModPhys.80.1355} {\bibfield  {journal}
  {\bibinfo  {journal} {Rev. Mod. Phys.}\ }\textbf {\bibinfo {volume} {80}},\
  \bibinfo {pages} {1355} (\bibinfo {year} {2008})}\BibitemShut {NoStop}%
\bibitem [{\citenamefont {{Farkov}}\ \emph {et~al.}(2019)\citenamefont
  {{Farkov}}, \citenamefont {{Manchanda}},\ and\ \citenamefont
  {{Siddiqi}}}]{Walsh}%
  \BibitemOpen
  \bibfield  {author} {\bibinfo {author} {\bibfnamefont {Y.~A.}\ \bibnamefont
  {{Farkov}}}, \bibinfo {author} {\bibfnamefont {P.}~\bibnamefont
  {{Manchanda}}},\ and\ \bibinfo {author} {\bibfnamefont {A.~H.}\ \bibnamefont
  {{Siddiqi}}},\ }\bibfield  {title} {\bibinfo {title} {{Introduction to Walsh
  Analysis and Wavelets}},\ }\href
  {https://doi.org//10.1007/978-981-13-6370-2_1} {\bibfield  {journal}
  {\bibinfo  {journal} {in, Construction of Wavelets Through Walsh Functions.
  Industrial and Applied Mathematics. Springer, Singapore}\ ,\ \bibinfo {pages}
  {1}} (\bibinfo {year} {2019})}\BibitemShut {NoStop}%
\bibitem [{\citenamefont {Magesan}\ and\ \citenamefont
  {Gambetta}(2020)}]{PhysRevA.101.052308}%
  \BibitemOpen
  \bibfield  {author} {\bibinfo {author} {\bibfnamefont {E.}~\bibnamefont
  {Magesan}}\ and\ \bibinfo {author} {\bibfnamefont {J.~M.}\ \bibnamefont
  {Gambetta}},\ }\bibfield  {title} {\bibinfo {title} {{Effective Hamiltonian
  models of the cross-resonance gate}},\ }\href
  {https://doi.org/10.1103/PhysRevA.101.052308} {\bibfield  {journal} {\bibinfo
   {journal} {Phys. Rev. A}\ }\textbf {\bibinfo {volume} {101}},\ \bibinfo
  {pages} {052308} (\bibinfo {year} {2020})}\BibitemShut {NoStop}%
\bibitem [{\citenamefont {Sundaresan}\ \emph {et~al.}(2020)\citenamefont
  {Sundaresan}, \citenamefont {Lauer}, \citenamefont {Pritchett}, \citenamefont
  {Magesan}, \citenamefont {Jurcevic},\ and\ \citenamefont
  {Gambetta}}]{sundaresan2020reducing}%
  \BibitemOpen
  \bibfield  {author} {\bibinfo {author} {\bibfnamefont {N.}~\bibnamefont
  {Sundaresan}}, \bibinfo {author} {\bibfnamefont {I.}~\bibnamefont {Lauer}},
  \bibinfo {author} {\bibfnamefont {E.}~\bibnamefont {Pritchett}}, \bibinfo
  {author} {\bibfnamefont {E.}~\bibnamefont {Magesan}}, \bibinfo {author}
  {\bibfnamefont {P.}~\bibnamefont {Jurcevic}},\ and\ \bibinfo {author}
  {\bibfnamefont {J.~M.}\ \bibnamefont {Gambetta}},\ }\bibfield  {title}
  {\bibinfo {title} {{Reducing Unitary and Spectator Errors in Cross Resonance
  with Optimized Rotary Echoes}},\ }\href
  {https://doi.org/10.1103/PRXQuantum.1.020318} {\bibfield  {journal} {\bibinfo
   {journal} {PRX Quantum}\ }\textbf {\bibinfo {volume} {1}},\ \bibinfo {pages}
  {020318} (\bibinfo {year} {2020})}\BibitemShut {NoStop}%
\bibitem [{Sup()}]{Supplement}%
  \BibitemOpen
  \href@noop {} {\bibinfo {title} {Supplemental material}}\BibitemShut
  {NoStop}%
\bibitem [{\citenamefont {Andersen}\ \emph {et~al.}(2020)\citenamefont
  {Andersen}, \citenamefont {Remm}, \citenamefont {Lazar}, \citenamefont
  {Krinner}, \citenamefont {Lacroix}, \citenamefont {Norris}, \citenamefont
  {Gabureac}, \citenamefont {Eichler},\ and\ \citenamefont
  {Wallraff}}]{Wallraff_s7_2020}%
  \BibitemOpen
  \bibfield  {author} {\bibinfo {author} {\bibfnamefont {C.~K.}\ \bibnamefont
  {Andersen}}, \bibinfo {author} {\bibfnamefont {A.}~\bibnamefont {Remm}},
  \bibinfo {author} {\bibfnamefont {S.}~\bibnamefont {Lazar}}, \bibinfo
  {author} {\bibfnamefont {S.}~\bibnamefont {Krinner}}, \bibinfo {author}
  {\bibfnamefont {N.}~\bibnamefont {Lacroix}}, \bibinfo {author} {\bibfnamefont
  {G.~J.}\ \bibnamefont {Norris}}, \bibinfo {author} {\bibfnamefont
  {M.}~\bibnamefont {Gabureac}}, \bibinfo {author} {\bibfnamefont
  {C.}~\bibnamefont {Eichler}},\ and\ \bibinfo {author} {\bibfnamefont
  {A.}~\bibnamefont {Wallraff}},\ }\bibfield  {title} {\bibinfo {title}
  {Repeated quantum error detection in a surface code},\ }\href
  {https://doi.org/10.1038/s41567-020-0920-y} {\bibfield  {journal} {\bibinfo
  {journal} {Nature Physics}\ }\textbf {\bibinfo {volume} {16}},\ \bibinfo
  {pages} {875} (\bibinfo {year} {2020})}\BibitemShut {NoStop}%
\bibitem [{\citenamefont {Chamberland}\ \emph {et~al.}(2020)\citenamefont
  {Chamberland}, \citenamefont {Zhu}, \citenamefont {Yoder}, \citenamefont
  {Hertzberg},\ and\ \citenamefont {Cross}}]{chamberland2020}%
  \BibitemOpen
  \bibfield  {author} {\bibinfo {author} {\bibfnamefont {C.}~\bibnamefont
  {Chamberland}}, \bibinfo {author} {\bibfnamefont {G.}~\bibnamefont {Zhu}},
  \bibinfo {author} {\bibfnamefont {T.~J.}\ \bibnamefont {Yoder}}, \bibinfo
  {author} {\bibfnamefont {J.~B.}\ \bibnamefont {Hertzberg}},\ and\ \bibinfo
  {author} {\bibfnamefont {A.~W.}\ \bibnamefont {Cross}},\ }\bibfield  {title}
  {\bibinfo {title} {Topological and subsystem codes on low-degree graphs with
  flag qubits},\ }\href {https://doi.org/10.1103/PhysRevX.10.011022} {\bibfield
   {journal} {\bibinfo  {journal} {Phys. Rev. X}\ }\textbf {\bibinfo {volume}
  {10}},\ \bibinfo {pages} {011022} (\bibinfo {year} {2020})}\BibitemShut
  {NoStop}%
\bibitem [{\citenamefont {Kandala}\ \emph {et~al.}(2021)\citenamefont
  {Kandala}, \citenamefont {Wei}, \citenamefont {Srinivasan}, \citenamefont
  {Magesan}, \citenamefont {Carnevale}, \citenamefont {Keefe}, \citenamefont
  {Klaus}, \citenamefont {Dial},\ and\ \citenamefont {McKay}}]{kandala2021}%
  \BibitemOpen
  \bibfield  {author} {\bibinfo {author} {\bibfnamefont {A.}~\bibnamefont
  {Kandala}}, \bibinfo {author} {\bibfnamefont {K.~X.}\ \bibnamefont {Wei}},
  \bibinfo {author} {\bibfnamefont {S.}~\bibnamefont {Srinivasan}}, \bibinfo
  {author} {\bibfnamefont {E.}~\bibnamefont {Magesan}}, \bibinfo {author}
  {\bibfnamefont {S.}~\bibnamefont {Carnevale}}, \bibinfo {author}
  {\bibfnamefont {G.~A.}\ \bibnamefont {Keefe}}, \bibinfo {author}
  {\bibfnamefont {D.}~\bibnamefont {Klaus}}, \bibinfo {author} {\bibfnamefont
  {O.}~\bibnamefont {Dial}},\ and\ \bibinfo {author} {\bibfnamefont {D.~C.}\
  \bibnamefont {McKay}},\ }\bibfield  {title} {\bibinfo {title} {{Demonstration
  of a High-Fidelity cnot Gate for Fixed-Frequency Transmons with Engineered
  $ZZ$ Suppression}},\ }\href {https://doi.org/10.1103/PhysRevLett.127.130501}
  {\bibfield  {journal} {\bibinfo  {journal} {Phys. Rev. Lett.}\ }\textbf
  {\bibinfo {volume} {127}},\ \bibinfo {pages} {130501} (\bibinfo {year}
  {2021})}\BibitemShut {NoStop}%
\bibitem [{\citenamefont {Wei}\ \emph {et~al.}(2021)\citenamefont {Wei},
  \citenamefont {Magesan}, \citenamefont {Lauer}, \citenamefont {Srinivasan},
  \citenamefont {Bogorin}, \citenamefont {Carnevale}, \citenamefont {Keefe},
  \citenamefont {Kim}, \citenamefont {Klaus}, \citenamefont {Landers},
  \citenamefont {Sundaresan}, \citenamefont {Wang}, \citenamefont {Zhang},
  \citenamefont {Steffen}, \citenamefont {Dial}, \citenamefont {McKay},\ and\
  \citenamefont {Kandala}}]{wei2021quantum}%
  \BibitemOpen
  \bibfield  {author} {\bibinfo {author} {\bibfnamefont {K.~X.}\ \bibnamefont
  {Wei}}, \bibinfo {author} {\bibfnamefont {E.}~\bibnamefont {Magesan}},
  \bibinfo {author} {\bibfnamefont {I.}~\bibnamefont {Lauer}}, \bibinfo
  {author} {\bibfnamefont {S.}~\bibnamefont {Srinivasan}}, \bibinfo {author}
  {\bibfnamefont {D.~F.}\ \bibnamefont {Bogorin}}, \bibinfo {author}
  {\bibfnamefont {S.}~\bibnamefont {Carnevale}}, \bibinfo {author}
  {\bibfnamefont {G.~A.}\ \bibnamefont {Keefe}}, \bibinfo {author}
  {\bibfnamefont {Y.}~\bibnamefont {Kim}}, \bibinfo {author} {\bibfnamefont
  {D.}~\bibnamefont {Klaus}}, \bibinfo {author} {\bibfnamefont
  {W.}~\bibnamefont {Landers}}, \bibinfo {author} {\bibfnamefont
  {N.}~\bibnamefont {Sundaresan}}, \bibinfo {author} {\bibfnamefont
  {C.}~\bibnamefont {Wang}}, \bibinfo {author} {\bibfnamefont {E.~J.}\
  \bibnamefont {Zhang}}, \bibinfo {author} {\bibfnamefont {M.}~\bibnamefont
  {Steffen}}, \bibinfo {author} {\bibfnamefont {O.~E.}\ \bibnamefont {Dial}},
  \bibinfo {author} {\bibfnamefont {D.~C.}\ \bibnamefont {McKay}},\ and\
  \bibinfo {author} {\bibfnamefont {A.}~\bibnamefont {Kandala}},\ }\href@noop
  {} {\bibinfo {title} {Quantum crosstalk cancellation for fast entangling
  gates and improved multi-qubit performance}} (\bibinfo {year} {2021}),\
  \Eprint {https://arxiv.org/abs/2106.00675} {arXiv:2106.00675 [quant-ph]}
  \BibitemShut {NoStop}%
\bibitem [{\citenamefont
  {B\"orner}(2020)}]{borner_simon-dominik_classical_2020}%
  \BibitemOpen
  \bibfield  {author} {\bibinfo {author} {\bibfnamefont {S.-D.}\ \bibnamefont
  {B\"orner}},\ }\emph {\bibinfo {title} {{Classical Chaos in Transmon Qubit
  Arrays}}},\ \href@noop {} {\bibinfo {type} {{Bachelor Thesis}}},\ \bibinfo
  {school} {University of Cologne} (\bibinfo {year} {2020})\BibitemShut
  {NoStop}%
\bibitem [{\citenamefont {Oganesyan}\ and\ \citenamefont
  {Huse}(2007)}]{PhysRevB.75.155111}%
  \BibitemOpen
  \bibfield  {author} {\bibinfo {author} {\bibfnamefont {V.}~\bibnamefont
  {Oganesyan}}\ and\ \bibinfo {author} {\bibfnamefont {D.~A.}\ \bibnamefont
  {Huse}},\ }\bibfield  {title} {\bibinfo {title} {{Localization of interacting
  fermions at high temperature}},\ }\href
  {https://doi.org/10.1103/PhysRevB.75.155111} {\bibfield  {journal} {\bibinfo
  {journal} {Phys. Rev. B}\ }\textbf {\bibinfo {volume} {75}},\ \bibinfo
  {pages} {155111} (\bibinfo {year} {2007})}\BibitemShut {NoStop}%
\bibitem [{\citenamefont {Monteiro}\ \emph {et~al.}(2021)\citenamefont
  {Monteiro}, \citenamefont {Micklitz}, \citenamefont {Tezuka},\ and\
  \citenamefont {Altland}}]{PhysRevResearch.3.013023}%
  \BibitemOpen
  \bibfield  {author} {\bibinfo {author} {\bibfnamefont {F.}~\bibnamefont
  {Monteiro}}, \bibinfo {author} {\bibfnamefont {T.}~\bibnamefont {Micklitz}},
  \bibinfo {author} {\bibfnamefont {M.}~\bibnamefont {Tezuka}},\ and\ \bibinfo
  {author} {\bibfnamefont {A.}~\bibnamefont {Altland}},\ }\bibfield  {title}
  {\bibinfo {title} {Minimal model of many-body localization},\ }\href
  {https://doi.org/10.1103/PhysRevResearch.3.013023} {\bibfield  {journal}
  {\bibinfo  {journal} {Phys. Rev. Research}\ }\textbf {\bibinfo {volume}
  {3}},\ \bibinfo {pages} {013023} (\bibinfo {year} {2021})}\BibitemShut
  {NoStop}%
\bibitem [{\citenamefont {M\'ezard}\ and\ \citenamefont
  {Montanari}(2009)}]{mezard_information}%
  \BibitemOpen
  \bibfield  {author} {\bibinfo {author} {\bibfnamefont {M.}~\bibnamefont
  {M\'ezard}}\ and\ \bibinfo {author} {\bibfnamefont {A.}~\bibnamefont
  {Montanari}},\ }\href@noop {} {\emph {\bibinfo {title} {{Information,
  Physics, and Computation}}}}\ (\bibinfo  {publisher} {{Oxford University
  Press}},\ \bibinfo {year} {2009})\BibitemShut {NoStop}%
\bibitem [{\citenamefont {Sheldon}\ \emph {et~al.}(2016)\citenamefont
  {Sheldon}, \citenamefont {Magesan}, \citenamefont {Chow},\ and\ \citenamefont
  {Gambetta}}]{PhysRevA.93.060302}%
  \BibitemOpen
  \bibfield  {author} {\bibinfo {author} {\bibfnamefont {S.}~\bibnamefont
  {Sheldon}}, \bibinfo {author} {\bibfnamefont {E.}~\bibnamefont {Magesan}},
  \bibinfo {author} {\bibfnamefont {J.~M.}\ \bibnamefont {Chow}},\ and\
  \bibinfo {author} {\bibfnamefont {J.~M.}\ \bibnamefont {Gambetta}},\
  }\bibfield  {title} {\bibinfo {title} {Procedure for systematically tuning up
  cross-talk in the cross-resonance gate},\ }\href
  {https://doi.org/10.1103/PhysRevA.93.060302} {\bibfield  {journal} {\bibinfo
  {journal} {Phys. Rev. A}\ }\textbf {\bibinfo {volume} {93}},\ \bibinfo
  {pages} {060302} (\bibinfo {year} {2016})}\BibitemShut {NoStop}%
\bibitem [{\citenamefont {Barends}\ \emph {et~al.}(2014)\citenamefont
  {Barends}, \citenamefont {Kelly}, \citenamefont {Megrant}, \citenamefont
  {Veitia}, \citenamefont {Sank}, \citenamefont {Jeffrey}, \citenamefont
  {White}, \citenamefont {Mutus}, \citenamefont {Fowler}, \citenamefont
  {Campbell}, \citenamefont {Chen}, \citenamefont {Chen}, \citenamefont
  {Chiaro}, \citenamefont {Dunsworth}, \citenamefont {Neill}, \citenamefont
  {O'Malley}, \citenamefont {Roushan}, \citenamefont {Vainsencher},
  \citenamefont {Wenner}, \citenamefont {Korotkov}, \citenamefont {Cleland},\
  and\ \citenamefont {Martinis}}]{Barends2014}%
  \BibitemOpen
  \bibfield  {author} {\bibinfo {author} {\bibfnamefont {R.}~\bibnamefont
  {Barends}}, \bibinfo {author} {\bibfnamefont {J.}~\bibnamefont {Kelly}},
  \bibinfo {author} {\bibfnamefont {A.}~\bibnamefont {Megrant}}, \bibinfo
  {author} {\bibfnamefont {A.}~\bibnamefont {Veitia}}, \bibinfo {author}
  {\bibfnamefont {D.}~\bibnamefont {Sank}}, \bibinfo {author} {\bibfnamefont
  {E.}~\bibnamefont {Jeffrey}}, \bibinfo {author} {\bibfnamefont {T.~C.}\
  \bibnamefont {White}}, \bibinfo {author} {\bibfnamefont {J.}~\bibnamefont
  {Mutus}}, \bibinfo {author} {\bibfnamefont {A.~G.}\ \bibnamefont {Fowler}},
  \bibinfo {author} {\bibfnamefont {B.}~\bibnamefont {Campbell}}, \bibinfo
  {author} {\bibfnamefont {Y.}~\bibnamefont {Chen}}, \bibinfo {author}
  {\bibfnamefont {Z.}~\bibnamefont {Chen}}, \bibinfo {author} {\bibfnamefont
  {B.}~\bibnamefont {Chiaro}}, \bibinfo {author} {\bibfnamefont
  {A.}~\bibnamefont {Dunsworth}}, \bibinfo {author} {\bibfnamefont
  {C.}~\bibnamefont {Neill}}, \bibinfo {author} {\bibfnamefont
  {P.}~\bibnamefont {O'Malley}}, \bibinfo {author} {\bibfnamefont
  {P.}~\bibnamefont {Roushan}}, \bibinfo {author} {\bibfnamefont
  {A.}~\bibnamefont {Vainsencher}}, \bibinfo {author} {\bibfnamefont
  {J.}~\bibnamefont {Wenner}}, \bibinfo {author} {\bibfnamefont {A.~N.}\
  \bibnamefont {Korotkov}}, \bibinfo {author} {\bibfnamefont {A.~N.}\
  \bibnamefont {Cleland}},\ and\ \bibinfo {author} {\bibfnamefont {J.~M.}\
  \bibnamefont {Martinis}},\ }\bibfield  {title} {\bibinfo {title}
  {Superconducting quantum circuits at the surface code threshold for fault
  tolerance},\ }\href {https://doi.org/10.1038/nature13171} {\bibfield
  {journal} {\bibinfo  {journal} {Nature}\ }\textbf {\bibinfo {volume} {508}},\
  \bibinfo {pages} {500} (\bibinfo {year} {2014})}\BibitemShut {NoStop}%
\bibitem [{\citenamefont {Yan}\ \emph {et~al.}(2018)\citenamefont {Yan},
  \citenamefont {Krantz}, \citenamefont {Sung}, \citenamefont {Kjaergaard},
  \citenamefont {Campbell}, \citenamefont {Orlando}, \citenamefont
  {Gustavsson},\ and\ \citenamefont {Oliver}}]{PhysRevApplied.10.054062}%
  \BibitemOpen
  \bibfield  {author} {\bibinfo {author} {\bibfnamefont {F.}~\bibnamefont
  {Yan}}, \bibinfo {author} {\bibfnamefont {P.}~\bibnamefont {Krantz}},
  \bibinfo {author} {\bibfnamefont {Y.}~\bibnamefont {Sung}}, \bibinfo {author}
  {\bibfnamefont {M.}~\bibnamefont {Kjaergaard}}, \bibinfo {author}
  {\bibfnamefont {D.~L.}\ \bibnamefont {Campbell}}, \bibinfo {author}
  {\bibfnamefont {T.~P.}\ \bibnamefont {Orlando}}, \bibinfo {author}
  {\bibfnamefont {S.}~\bibnamefont {Gustavsson}},\ and\ \bibinfo {author}
  {\bibfnamefont {W.~D.}\ \bibnamefont {Oliver}},\ }\bibfield  {title}
  {\bibinfo {title} {Tunable coupling scheme for implementing high-fidelity
  two-qubit gates},\ }\href {https://doi.org/10.1103/PhysRevApplied.10.054062}
  {\bibfield  {journal} {\bibinfo  {journal} {Phys. Rev. Applied}\ }\textbf
  {\bibinfo {volume} {10}},\ \bibinfo {pages} {054062} (\bibinfo {year}
  {2018})}\BibitemShut {NoStop}%
\bibitem [{\citenamefont {Neill}(2017)}]{Neill2018}%
  \BibitemOpen
  \bibfield  {author} {\bibinfo {author} {\bibfnamefont {C.}~\bibnamefont
  {Neill}},\ }\emph {\bibinfo {title} {{A path towards quantum supremacy with
  superconducting qubits}}},\ \href
  {https://web.physics.ucsb.edu/~martinisgroup/theses/Neill2017.pdf} {\bibinfo
  {type} {{PhD Thesis}}},\ \bibinfo  {school} {University of California}
  (\bibinfo {year} {2017})\BibitemShut {NoStop}%
\bibitem [{foo()}]{footnorm}%
  \BibitemOpen
  \href@noop {} {\bibinfo {title} {{Here is the procedure for creating the
  specific disorder realization for variable $\delta E_J$ for our $3\times 3$
  array: Draw nine independent values $v_i$ ($1\leq i\leq 9$) from the standard
  normal distribution $\mathcal{N}(0,1)$. Then the $E_J$ value at site $i$ is
  $E_{Ji}=12.58{\mbox{MHz}}+v_i\delta E_J$ if site $i$ is $A$ type, and
  $E_{Ji}=13.80{\mbox{MHz}}+v_i\delta E_J$ if site $i$ is $B$
  type.}}}\BibitemShut {Stop}%
\end{thebibliography}%

\end{document}